\documentclass{article}

\usepackage{amsmath,marvosym,fullpage,bm}
\usepackage{psfrag}             
\usepackage{subfigure}          
\usepackage{graphicx}
\usepackage{wrapfig}
\usepackage{amsfonts}
\usepackage{amssymb}

\begin{document}

\title{A quasi-physical dynamic reduced order model for thermospheric mass density via Hermitian Space Dynamic Mode Decomposition}


\author{
Piyush M. Mehta\thanks{Research Associate, Department of Aerospace Engineering and Mechanics, Email: piyushmukeshmehta@gmail.com.},\
Richard Linares\thanks{Assistant Professor, Department of Aerospace Engineering and Mechanics. Email: rlinares@umn.edu.}
\\
{\normalsize\itshape
    University of Minnesota, Minneapolis, NM, USA}\\
Eric K. Sutton\thanks{Research Scientist}
\\
{\normalsize\itshape
    Air Force Research Laboratory, Kirtland AFB, Albuquerque, New Mexico, USA}\\
}




%
%
%
%
%


\date{}

\maketitle

\begin{abstract}
Thermospheric mass density is a major driver of satellite drag, the largest source of uncertainty in accurately predicting the orbit of satellites in low Earth orbit (LEO) pertinent to space situational awareness. Most existing models for thermosphere are either physics-based or empirical. Physics-based models offer the potential for good predictive/forecast capabilities but require dedicated parallel resources for real-time evaluation and data assimilative capabilities that have yet to be developed. Empirical models are fast to evaluate, but offer very limited forecasting abilities. This paper presents methodology for developing a reduced-order dynamic model from high-dimensional physics-based models by capturing the underlying dynamical behavior. The quasi-physical reduced order model (ROM) for thermospheric mass density is developed using a large dataset of TIE-GCM (Thermosphere-Ionosphere-Electrodynamics General Circular Model) simulations spanning 12 years and covering a complete solar cycle. Towards this end, a new reduced order modeling approach, based on Dynamic Mode Decomposition with control (DMDc), that uses the Hermitian space of the problem to derive the dynamics and input matrices in a tractable manner is developed. Results show that the ROM performs well in serving as a reduced order surrogate for TIE-GCM while almost always maintaining the forecast error to within 5\% of the simulated densities after 24 hours.
\end{abstract}


\section{Introduction}  

Space situational awareness (SSA) has come into the spotlight in recent years because of the increasing dependence of mankind on space assets and the large number of debris objects that constantly endanger them. The plans to inject thousands of additional satellites as part of multiple technology-advancing mega-constellations into orbits already densely populated puts further stress on SSA operations (\cite{MegaC}). In the event of a collision, the smallest of debris objects, traveling at relative velocities of several km/sec, can render significant damage to systems onboard the space assets. In addition, each collision event can push the space environment closer to a cascade, otherwise known as the Kessler syndrome \cite{Kessler}, that can render space itself inaccessible for future generations. Mitigating this threat requires the integration of comprehensive SSA into operations.

In low Earth orbit (LEO), one of the most densely populated orbital regimes, atmospheric drag is the largest source of uncertainty in our ability to accurately predict the orbit of resident space objects (RSOs). Thermospheric mass density, which feeds directly into the drag model, is one of the major causes behind the uncertainty. The thermosphere is a highly dynamic environment with the Sun being the strongest driver of its variations. The thermosphere is readily heated by the Sun's irradiance, especially in the EUV spectrum, causing significant variations of mass density. Other sources of variations include solar rotation and cycle, diurnal and higher order harmonics, magnetic storms and substorm, gravity waves, winds and tides, and long term climate change. A review of the basic dynamics and drivers is given by \cite{Forbes}. A review of thermospheric mass density behavior in the context of satellite drag is provided by \cite{Emmert}. 

Existing models for thermospheric mass density can be classified either as physics-based or empirical. Physics-based models are based on first principles and solve the fluid equations by discretizing over the volume of interest resulting in hundreds of thousands of solved for states. Solving the system of equations dynamically provides the models with good predictive/forecasting capabilities; however, the true predictive/forecasting potential cannot be unlocked until the data assimilation schemes are improved. On the other end of the spectrum, empirical models specify the average behavior of the thermosphere with parameterized functions formulated using measurements or observations from multiple sources. Empirical models are fast to evaluate due to their simplified mathematical formulation; however, they offer very limited forecasting abilities since they do not model the system dynamically. Since either type of models can have large biases and/or errors, data assimilation is almost always required and is a very active area of research.

This work develops methodology for a quasi-physical dynamic reduced order model (ROM) of the thermosphere that has the desired properties of both the physics-based and empirical models, and can be easily and readily integrated into SSA operations. A reduced order model facilitates real-time large ensemble evaluations for improved characterization of model uncertainty and collision probabilities at a significantly reduced cost. The approach can also provide insights into the model errors associated with the underlying dynamics, which will be a subject of future work. The new method uses a dynamic systems formulation that inherently provides predictive capabilities, an advantage and improvement over previous work using the Proper Orthogonal Decomposition (POD) or Empirical Orthogonal Function (EOF) approach (\cite{Mehta_POD,EOF1}). Observations can be introduced into the model through data assimilation, which will also be a subject of future work.

Modal decomposition or variance reductions methods, highly popular in the fluid dynamics community (\cite{MA_review}), provide a path to achieving an efficient method for estimating the state of the thermosphere. The main idea behind modal decomposition is that a large fraction of the total energy/variance of a system can be captured through projection along a handful of dominant directions or dimensions, referred to in this work as modes. The modes represent time-independent coherent structures that exist in the flow, which when combined with time-dependent coefficients can reproduce the total energy/variance of the system.

\cite{MA_review} provide an overview of existing modal decomposition methods applied to fluid flows. They classify the method as data-based or operator-based. Data-based approaches that include proper orthogonal decomposition (POD) and Dynamic Mode Decomposition (DMD) are ideal for cases where high-fidelity measurements are available with little to no knowledge of the system dynamics. Operator-based approaches that include the well-known Koopman analysis require knowledge of the system dynamics for order reduction. Even though we may have knowledge of the dynamic equations behind physics-based models, we choose a data-based approach to keep the development completely data-driven and for the ease of implementation.

\cite{Mehta_POD} presented a methodology for reduced order modeling and calibration of the upper atmosphere based on proper orthogonal decomposition (POD). They developed the methodology using the Naval Research Laboratory's MSIS model. Although MSIS is already a highly simplified implementation with small computational expense, it was an ideal platform for testing the methodology. The method was able to accurately identify the known dynamics in the MSIS model as POD modes or basis functions while providing an approach for calibration of the model. For more details, the reader is directed to \cite{Mehta_POD}. 

The POD approach, also known as Empirical Orthogonal Functions (EOFs), has been previously applied to infer thermosphere dynamics from discrete measurements along the satellite's path [\cite{EOF1}, \cite{EOF2}]; however, the work has been restricted to 2-dimensional structures, typically at a constant altitude and does not constitute a model, per se. EOF based analysis has also been used by Sutton et al. (2012) to extract modes of the thermosphere from a physics-based model and use them as a replacement for the generic spherical harmonic modes commonly used in semi-empirical models. The methodology developed by \cite{Mehta_POD} is 3-dimensional and fully data-driven, i.e. the 3-dimensional coherent structures of the modes are derived using simulation output from a model over the volumetric grid of interest, and constitutes a model. 

In this paper, we seek to develop a ROM for physics-based models that represent dynamic systems with exogenous inputs. Because model order reduction for the upper atmosphere requires simulation data covering a full solar cycle lasting over a decade, we develop a new method to keep the problem tractable. The new method is based on Dynamic Mode Decomposition with control (DMDc) (\cite{DMDc}), and uses the Hermitian Space of the problem towards a comprehensive quasi-physical dynamic ROM for thermospheric mass density using a dynamic system formulation that has not been attempted before. We call the new method Hermitian Space - Dynamic Mode Decomposition with control (HS-DMDc). The dynamic system formulation provides inherent predictive capabilities while significantly simplifying the process of assimilating data. We develop a proof of concept quasi-physical dynamic ROM for NCAR's Thermosphere-Ionosphere-Electrodynamics General Circular Model (TIE-GCM).

The paper is structured as follows: section \ref{s:Methods} describes the data-based variance reduction methods including POD, DMD and the newly developed HS-DMDc method. Section \ref{s:MD} provides details about the process of model development including method validation in section \ref{s:MeV} using yearly simulations for 2002 and 2009; developing a universally applicable model in section \ref{s:UM} using 12 years of TIE-GCM simulations covering a full solar cycle; and validation of the universal model in section \ref{s:UMV}. Section \ref{s:ML} discusses the model limitations with section \ref{s:FD} describing further development steps. Finally, section \ref{s:Conc} concludes the paper.  

\section{Methodology}\label{s:Methods}
Data-driven, equation-free methods for order reduction of complex high-dimensional systems have become very popular in the fluids community (\cite{MA_review}). Most decomposition methods were originally developed as diagnostic tools for characterizing complex fluid flows by extracting physically interpretable spatial structures or modes and their associated temporal responses. The POD technique, first introduced to the fluids community by \cite{Lumley}, serves as the basis and motivation for the development of other modal decomposition techniques, including DMD, DMDc, and HS-DMDc. In this work, we will limit our discussion on modal decomposition techniques to POD, DMD, and the derivation of HS-DMDc. Readers are referred to \cite{DMDc} for details about DMDc.

\subsection{Proper Orthogonal Decomposition}\label{s:POD}
The two most commonly used formalisms of POD are the method of snapshots and singular value decomposition (SVD). The idea behind POD is the derivation of an optimal set of basis functions or modes for a given set of snapshots. Snaphots can be a collection of the model/simulation output or experimental data acquired over time. The POD framework seeks to split space and time by decomposing the energy/variance of a spatially and temporally varying field. Both formalisms provide an optimal set of orthogonal vectors; however, they differ in the manner in which the modes are derived.  The method of snapshots, used in \cite{Mehta_POD}, decomposes the variance after removing the mean component in the following manner
\begin{equation}\label{e:POD1}
\tilde{\textbf{x}}({\bf s},{ t}) = \textbf{x}({\bf s},{t}) - \bar{\textbf{x}}({\bf s}) \approx \sum_{i=1}^{r} {c}_i({ t})\pmb{\phi}_{i}({\bf s}) 
\end{equation}
where $\textbf{x}({\bf s},{ t})$ is a random field (in this case the neutral mass density) on a spatial domain (in this case a uniform grid in local time, latitude and altitude), $\bar{\textbf{x}}({\bf s})$ is the temporal mean, $\tilde{\textbf{x}}({\bf s},{t})$ is the variance, ${\bf s}$ is the spatial vector (number of spatial points saved per time snapshot unfolded into a column vector of size $n$), and $t$ is the time. The POD modes, $\pmb{\phi}_{i}({\bf s})$, are purely spatial while the temporal response is given by the coefficients, ${c}_{i}({t})$, where $r$ is the number of modes used to construct the truncated low order representation. The POD modes are derived using an eigendecomposition of the square correlation matrix (using the Hermitian Space) that represents the distance between the snapshots. For details about application of the POD method of snapshots to reduced order modeling of neutral mass density, the reader is referred to \cite{Mehta_POD}.

The SVD can be thought of as a generalization of the eigendecomposition (used in method of snapshots) to rectangular matrices. SVD decomposes a rectangular matrix ${\bf M} \in \mathbb{R}^{n \times m}$ as
\begin{equation}\label{e:SVD}
{\bf M}={\bf U} \mathbf{\Sigma} {\bf V}^T
\end{equation}
where ${\bf U} \in \mathbb{C}^{n \times n}$ and ${\bf V} \in \mathbb{C}^{m \times m}$ are unitary matrices, and $\mathbf{\Sigma} \in \mathbb{R}^{n \times m}$ is a diagonal matrix. The left singular vectors ${\bf U}$ that span the range of ${\bf M}$ are orthogonal and optimal, and represent the POD modes. In practice, one only needs to compute a reduced SVD correponding to the non-zero singular values, henceforth referred to in this work as economy SVD (E-SVD), of which there are at most $min(n,m)$. The SVD formalism decomposes the snapshots directly (${\bf M}$ contains a series of snapshots) without taking away the mean component; and therefore, the first mode or singular vector of ${\bf U}$ contains a strong mean component. The eigenvalue and singular value decomposition are closely related; the left singular vectors, ${\bf U}$, represent the eigenvectors of ${\bf M}{\bf M}^{T}$ (the correlation matrix decomposed in the method of snapshots before taking away the mean), where $T$ denotes the conjugate transpose. Therefore, if the SVD decomposition is performed on the snapshots after subtracting the mean component, the POD modes from method of snapshots and SVD will be equivalent. For more details about eigenvalue and singular value decomposition, the reader if referred to \cite{MA_review}.

\subsection{Dynamic Mode Decomposition}\label{s:DMD}
The SVD formalism of POD sits at the heart of DMD. The weakness of POD that DMD seeks to overcome is that the temporal coefficients of the POD modes generally contain a mix of frequencies and does not allow a formulation for forecast or prediction. \cite{Mehta_POD} overcame this using fast Fourier transform (FFT) coupled with a Gaussian Process model for the coefficients. DMD can be thought of as an ideal combination of POD with Fourier transforms in time, resulting in the DMD modes associated with a single frequency with a possible growth or decay rate. The idea behind DMD is the derivation of a best-fit linear dynamical system by fitting the time domain data or snapshots obtained from the underlying nonlinear dynamical system. Consider a continuous time dynamical system given as
\begin{equation}\label{e:DS1}
\frac{d{\bf x}}{dt}=F({\bf x},t;\mathbf{\Theta})
\end{equation}
where $F(\cdot)$ represents the dynamics, $\mathbf{\Theta}$ is the set of system parameters, and ${\bf x}$ is the state vector $\in \mathbb{R}^{n}$ (comprising of the random field on the spatial vector ${\bf s}$). Because a closed form solution for the evolution of the dynamic system is generally not feasible, a numerical solution approach is typically used, as in the case of the physics-based models. DMD uses an equation-free approach (not operating on the physical dynamic equations) to construct an approximate locally linear dynamical system
\begin{equation}\label{e:DS2}
\frac{d{\bf x}}{dt}=\pmb{\mathcal{A}}{\bf x}
\end{equation}
under the assumption that the linear operator $\pmb{\mathcal{A}}$ is diagonalizable. Given the initial condition ${\bf x}(0)$, the above system has a well known solution [\cite{EDE}]
\begin{equation}\label{e:DS3}
{\bf x}(t)=\sum_{i=1}^{n}b_i \exp(\omega_i t)\pmb{\phi}_{i} = {\bf b}\exp(\mathbf{\Omega}t)\mathbf{\Phi}
\end{equation}
where $\omega_i$ and $\phi_i$ are eigenvalues and eigenvectors of the dynamic matrix $\pmb{\mathcal{A}}$, and $b_i$ are the coefficients corresponding to the initial condition in the eigenvector basis.
Since the snapshots are samples from the continuous system sampled in time ($\Delta t$), an analogous discrete-time system is given by
\begin{equation}\label{e:DS4}
{\bf x}_{k+1}={\bf A}{\bf x}_k
\end{equation}
where ${\bf A} \in \mathbb{R}^{n \times n}$ is the discrete time map of $\pmb{\mathcal{A}}$ such that
\begin{equation}\label{e:DS5}
{\bf A}=\exp(\pmb{\mathcal{A}}\Delta t)
\end{equation}
The solution to the discrete system can be given with the eigenvalues ($\lambda$) and eigenvectors ($\pmb{\phi}$) of ${\bf A}$ as 
\begin{equation}\label{e:DS6}
{\bf x}_k \approx \sum_{i=1}^{r}b_i \lambda_i^k \pmb{\phi}_i = {\bf b}\mathbf{\Lambda}^k\mathbf{\Phi}
\end{equation}
where ${\bf b}$ again are the initial conditions coefficients in the eigenvector basis. The eigenvector basis ($\pmb{\phi}$) in Eq. \ref{e:DS6} represents dynamic modes that differ from the modes derived in POD and are not orthogonal. The low order eigendecomposition of the matrix ${\bf A}$ produced by DMD represents an optimal fit to the measured trajectory in the least squares sense, minimizing $\| {\bf x}_{k+1} - {\bf A}{\bf x}_k \|$ across all snapshots. The above description of DMD is derived from \cite{DMD_book}, where the readers can find further details about the method.

\subsubsection{DMD Algorithm}
The DMD algorithm provides a solution to the discrete-time system in Eq.~\ref{e:DS4} by extracting an estimate of the dynamic matrix ${\bf A}$ by rearranging a series of outputs from a dynamical system or snapshots, ${\bf x}_k$, into time-shifted data matrices. Let ${\bf X}_1$ and ${\bf X}_2$ be the time-shifted matrix of snapshots such that
\begin{equation}\label{e:DMD1}
{\bf X}_1 = \left[ {\bf x}_1, \quad {\bf x}_2, \quad \cdots, {\bf x}_{m-1}\right], \quad \quad
{\bf X}_2 = \left[ {\bf x}_2, \quad {\bf x}_3, \quad \cdots, {\bf x}_{m}\right]
\end{equation}
where $m$ is the number of snapshots. The data matrices ${\bf X}_1$ and ${\bf X}_2$ can be related (${\bf X}_2$ is the time evolution of ${\bf X}_1$) through a best-fit linear model as in Eq.~\ref{e:DS4} such that
\begin{equation}\label{e:DMD2}
{\bf X}_2 = {\bf A}{\bf X}_1. 
\end{equation}
The dynamic matrix ${\bf A}$ is estimated as ${\bf A}={\bf X}_2{\bf X}_1^{\dagger}$, where ${\bf X}_1^{\dagger}$ represents the pseudoinverse of the snapshot matrix ${\bf X}_1$. The pseudoinverse is calculated using a E-SVD such that ${\bf X}_1 = {{\bf U}_{r}}\mathbf{\Sigma}_{r} {\bf V}_{r}^T$ and ${\bf X}_1^{\dagger} = {\bf V}_{r}\mathbf{\Sigma}_{r}^{-1}{\bf U}_{r}^T$, where $r$ is the reduced rank. Since computing and storing the full order dynamic matrix, ${\bf A} \in \mathbb{R}^{n \times n}$, can be computationally infeasible when $n \gg 1$, a reduced order approximation of the dynamic matrix, $\tilde{{\bf A}} \in \mathbb{R}^{r \times r}$, is derived using a similarity transform with projection onto a reduced set of orthogonal basis vectors. Using a reduced set of the left singular vectors ${\bf U}_{r} \in \mathbb{R}^{n \times r}$ from the E-SVD of ${\bf X}_1$, we get
\begin{equation}\label{e:DMD3}
{\bf z}_k = {\bf U}_{r}^{\dagger}{\bf x}_k = {\bf U}_{r}^{T}{\bf x}_k. 
\end{equation}
where ${\bf z} \in \mathbb{R}^{r}$ is the reduced order state vector. Substituting Eq.~\ref{e:DMD3} into Eq.~\ref{e:DS4} gives
\begin{equation}\label{e:DMD4}
{\bf U}_{r}{\bf z}_{k+1} = {\bf A}{\bf U}_{r}{\bf z}_{k} 
\end{equation}
Now, multiplying both sides of the above equation by ${\bf U}^{\dagger}$, we get
\begin{equation}\label{e:DMD5}
{\bf z}_{k+1} = {\bf U}_{r}^{\dagger}{\bf A}{\bf U}_{r}{\bf z}_{k} = {\bf U}_{r}^{T}{\bf A}{\bf U}_{r}{\bf z}_{k} = \tilde{{\bf A}}{\bf z}_{k}
\end{equation}
The reduced order state vector ${\bf z}$ represents the coefficients corresponding to the left singular vectors or POD modes.
The algorithm steps are given below:
\begin{enumerate}
	\item Construct the snapshot matrices ${\bf X}_1$ and ${\bf X}_2$. 
	\item Perform E-SVD ${\bf X}_1 = {\bf U}_{r}\mathbf{\Sigma}_{r} {\bf V}_{r}^T$ to compute the psuedoinverse ${\bf X}_1^{\dagger}= {\bf V}_{r}\mathbf{\Sigma}_{r}^{-1}{\bf U}_{r}^T$ . 
	\item Compute the reduced order dynamic matrix $\tilde{{\bf A}} = {\bf U}_{r}^T{\bf A}{\bf U}_{r} = {\bf U}_{r}^T{\bf X}_2{\bf V}_{r}\mathbf{\Sigma}_{r}^{-1}$
	\item Compute the DMD modes as	$\mathbf{\Phi}={\bf U_{r}}{\bf W}$, where ${\bf W}$ are the eigenvectors of $\tilde{{\bf A}}$ such that $\tilde{{\bf A}}{\bf W}={\bf W}\mathbf{\Lambda}$.
\end{enumerate}

\subsection{Hermitian Space - Dynamic Mode Decomposition with control}\label{s:NM}
The HS-DMDc method developed here is an extension of DMD to dynamical systems with exogenous inputs. The method draws inspiration from the equation-free Dynamic Mode Decomposition with control (DMDc) algorithm derived by \cite{DMDc} that also builds on DMD, but can extract both the underlying dynamics and the input-output characteristics of a dynamical system. The method can be used to construct a ROM of the high-dimensional system for future state prediction under the influence of dynamics and external control. Unlike DMD, the snapshots include the state and input(s). The method characterizes the relationship between the future state, ${\bf x}_{k+1}$, the current state, ${\bf x}_k$, and the current input, ${\bf u}_k$, with a locally linearized model
\begin{equation}\label{e:DMDc1}
{\bf x}_{k+1} = {\bf A}{\bf x}_k + {\bf B}{\bf u}_k
\end{equation}
where ${\bf x} \in \mathbb{R}^{n}$, ${\bf u} \in \mathbb{R}^{q}$, ${\bf A} \in \mathbb{R}^{n \times n}$, and ${\bf B} \in \mathbb{R}^{n \times q}$. The dynamic matrix ${\bf A}$ describes the unforced dynamics of the system and the input matrix ${\bf B}$ characterizes the effect of input ${\bf u}_k$ on the state ${\bf x}_{k+1}$. 

The difference between the HS-DMDc and DMDc algorithms is the formalism used in the computation of the pseudoinverse and the left singular vectors. In order for the developed model (estimates of the dynamic and input matrices, $\bf A$ and $\bf B$) to be applicable for all space weather conditions, the simulated snapshots need to represent the full range of inputs. Because the solar cycle lasts over a decade, this requires a large data set of more than ($m \approx$) 400,000 snapshots with a 0.25 hr resolution. A 5 degree grid resolution in TIE-GCM results in a state vector size of ($n \approx$) 75,000 with a 2.5 degree grid resolution resulting in $n \approx 300,000$.

Large data has motivated extensions to DMD even beyond E-SVD (\cite{Hemati,Erichson}), but have been limited to systems with no exogenous inputs. The theoretical computational complexity of full rank SVD of ${\bf X}_1 \in \mathbb{R}^{n \times m}$ used in DMDc is $O(mn^2)$ with $n\leq m$, making its application intractable for the problem at hand. The use of E-SVD reduces the complexity to $O(mnr)$ (\cite{SVD_Lit}) by computing only the first $r$ singular values and vectors. HS-DMDc reduces the computation of the psuedoinverse ($^\dagger$) to the Hermitian space by performing an eigendecomposition of the correlation matrix, ${\bf X}_1 {\bf X}_1^T \in \mathbb{R}^{n \times n}$, reducing the full  rank complexity to $O(nn^2)$. The complexity can be reduced to $O(n^2r)$ using an economy EigenDecomposition (E-ED). In theory, the computation of the correlation matrix ${\bf X}_1 {\bf X}_1^T$ also introduces linear scaling with $m$ - $O(mn^2)$. Although formulating the problem in the Hermitian space is somewhat of a common practice, motivated in part by the method of snapshot formalism of POD, it is important to note that using Eigendecomposition to compute the singular values and vectors can be more sensitive to numerical roundoff errors.

\begin{figure}[h]
	\centering
	\includegraphics[width=0.7\textwidth]{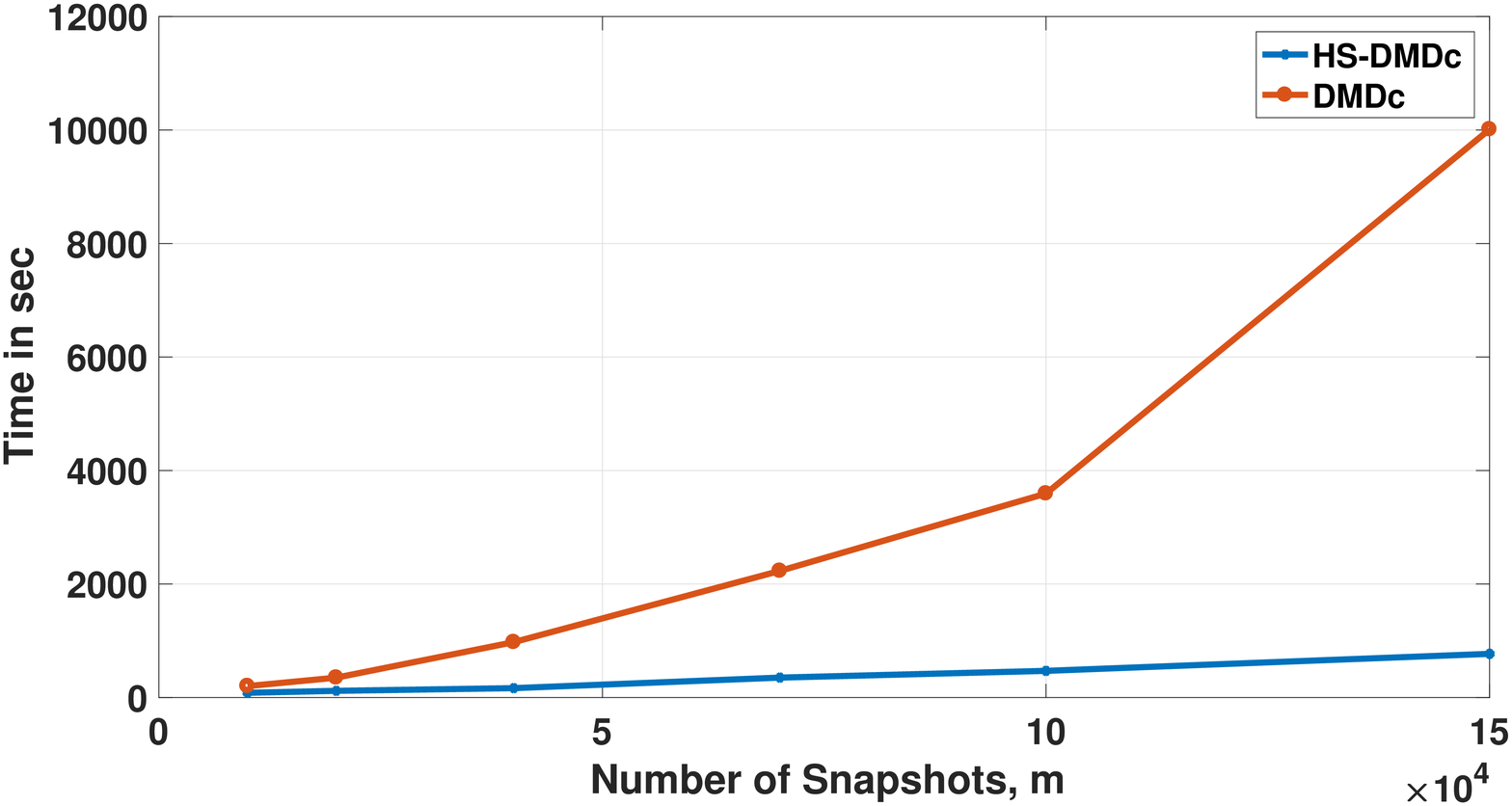}
	\caption{Numerical computational cost comparison between HS-DMDc and DMDc.}
	\label{f:Numerical_Cost}
\end{figure}

Because in practice the cost depends on several factors, we perform a simple representative numerical study to highlight the cost savings from HS-DMDc. The study is performed using Matlab$\textsuperscript{\textregistered}$ on a Macbook Pro: 3.1GHz Intel i7 with 16GB of RAM. The study uses the $svds$ and $eigs$ functions to compute the first 20 most energetic POD modes for state size's $n = $ 10,000; the number of snapshots $m$ is varied from 10,000 to 150,000. The computational cost comparison is shown in Figure \ref{f:Numerical_Cost}. In this case, HS-DMDc offers cost savings of close to an order or magnitude for $m = 100,000$ with the savings growing to two orders of magnitude for $m = 150,000$. We expect similar, if not better savings for increasing $n$ and/or $r$.

HS-DMDc uses the same time-shifted snapshot matrices as defined for DMD; however, because the system now includes external control defined as $\mathbf{\Upsilon}=[{\bf u}_1,\;{\bf u}_2,\; \cdots,\; {\bf u}_{m-1}]$, Eq.~\ref{e:DMD2} is modified such that 
\begin{equation}\label{e:DMDc2}
{\bf X}_2 = {\bf Z}\mathbf{\Psi}
\end{equation}
where ${\bf Z}$ and $\mathbf{\Psi}$ are the augmented operator and data matrices respectively.
\begin{equation}\label{e:DMDc3}
{\bf Z} \triangleq \begin{bmatrix} {\bf A} & {\bf B}
\end{bmatrix} \quad \text{and} \quad \mathbf{\Psi} \triangleq \begin{bmatrix} {\bf X}_1 \\ 
\mathbf{\Upsilon}
\end{bmatrix}
\end{equation}
The goal again is to estimate the dynamic and input matrices while minimizing $\| {\bf X}_{2} - {\bf Z}\mathbf{\Psi} \|$. The augmented operator matrix is solved for just as in DMD, ${\bf Z}={\bf X}_2\mathbf{\Psi}^{\dagger}$; however, the psuedoinverse of $\mathbf{\Psi}$ is computed in the Hermitian space using E-ED such that $\mathbf{\Psi}^{\dagger}=\mathbf{\Psi}^T(\mathbf{\Psi}\mathbf{\Psi}^T)^{-1}$ and $(\mathbf{\Psi}\mathbf{\Psi}^T)^{-1}=(\hat{{\bf U}}_{\hat{r}}\hat{\mathbf{\Xi}}_{\hat{r}}\hat{{\bf U}}_{\hat{r}}^{\dagger})^{-1}=\hat{{\bf U}}_{\hat{r}}\hat{\mathbf{\Xi}}_{\hat{r}}^{-1}\hat{{\bf U}}_{\hat{r}}^T$. The orthogonal basis vectors $\hat{{\bf U}}_{\hat{r}} \in \mathbb{R}^{(n+p) \times \hat{r}}$, equivalent to the left singular vectors of a SVD of $\mathbf{\Psi}$, are eigenvectors of the correlation matrix $\mathbf{\Psi}\mathbf{\Psi}^T$ (such that $\mathbf{\Psi}\mathbf{\Psi}^T\hat{{\bf U}}_{\hat{r}}=\hat{{\bf U}}_{\hat{r}}\hat{\mathbf{\Xi}}_{\hat{r}}$), where $p$ is the number of inputs and $\hat{r}$ is the low rank truncation value for E-ED of $\mathbf{\Psi}$.

The dynamic and input matrices can then be estimated as 
\begin{equation}\label{e:DMDc4}
{\bf A} = {\bf X}_2\mathbf{\Psi}\hat{{\bf U}}_{\hat{r}}\hat{\mathbf{\Xi}}_{\hat{r}}^{-1}\hat{{\bf U}}_{\hat{r},1}^T \qquad \text{and} \qquad {\bf B} = {\bf X}_2\mathbf{\Psi}\hat{{\bf U}}_{\hat{r}}\hat{\mathbf{\Xi}}_{\hat{r}}^{-1}\hat{{\bf U}}_{\hat{r},2}^T
\end{equation}
where $\hat{\mathbf{\Xi}}_{\hat{r}} \in \mathbb{R}^{\hat{r} \times \hat{r}}$ are the eigenvalues and $\hat{{\bf U}}_{\hat{r}}^T = [\hat{{\bf U}}_{\hat{r},1}^T \; \hat{{\bf U}}_{\hat{r},2}^T]$ with $\hat{{\bf U}}_{\hat{r},1} \in \mathbb{R}^{n \times \hat{r}}$ and $\hat{{\bf U}}_{\hat{r},2} \in \mathbb{R}^{p \times \hat{r}}$. Again, the reduced order or low rank approximations for the dynamic and input matrices are achieved through projection onto a truncated POD basis. This however, requires an additional E-ED in the Hermitian space for either ${\bf X}_1$ or ${\bf X}_2$ since $\hat{{\bf U}}_{\hat{r}}$ is defined in the input space and projection is performed in the output space. Substituting Eq.~\ref{e:DMD3} into Eq.~\ref{e:DMDc1}, we get
\begin{equation}\label{e:DMDc5}
{\bf U}_{r}{\bf z}_{k+1} = {\bf A}{\bf U}_{r}{\bf z}_{k} + {\bf B}{\bf u}_k
\end{equation}
where $\mathbf{U}_{r} \in \mathbb{R}^{n \times r}$ are the orthogonal eigenvectors such that ${\bf X}_1{\bf X}_1^T{\bf U}_{r}={\bf U}_{r}\mathbf{\Xi}_{r}$, and $r$ is the low rank truncation value such that $\hat{r} > r$. Multiplying both sides by ${\bf U}_{r}^{\dagger}$, we get
\begin{equation}\label{e:DMDc6}
{\bf z}_{k+1} = {\bf U}_{r}^{\dagger}{\bf A}{\bf U}_{r}{\bf z}_{k} + {\bf U}_{r}^{\dagger}{\bf B}{\bf u}_k
= \tilde{{\bf A}}{\bf z}_{k} + \tilde{{\bf B}}{\bf u}_k 
\end{equation}
The reduced order state vector again represents the coefficients of the POD modes. The reduced order approximations for the dynamic and input matrices are then computed as
\begin{equation}\label{e:DMDc7}
\tilde{{\bf A}} = {\bf U}_{r}^T{\bf A}{\bf U}_{r} = {\bf U}_{r}^T{\bf X}_2\mathbf{\Psi}\hat{{\bf U}}_{\hat{r}}\hat{\mathbf{\Xi}}_{\hat{r}}^{-1}\hat{{\bf U}}_{\hat{r},1}^T{\bf U}_{r} \qquad \text{and}  \qquad \tilde{{\bf B}} = {\bf U}_{r}^T{\bf B} = {\bf U}_{r}^T{\bf X}_2\mathbf{\Psi}\hat{{\bf U}}_{\hat{r}}\hat{\mathbf{\Xi}}_{\hat{r}}^{-1}\hat{{\bf U}}_{\hat{r},2}^T
\end{equation}
where $\mathbf{\Xi}_{r} \in \mathbb{R}^{r \times r}$ are the eigenvalues.

\subsubsection{HS-DMDc Algorithm}\label{s:NA}
\begin{enumerate}
	\item Construct the data matrices ${\bf X}_1$, ${\bf X}_2$, $\mathbf{\Upsilon}$, and $\mathbf{\Psi}$.
	\item Perform E-ED in the Hermitian space, $\mathbf{\Psi}\mathbf{\Psi}^T=\hat{{\bf U}}_{\hat{r}}\hat{\mathbf{\Xi}}_{\hat{r}}\hat{{\bf U}}_{\hat{r}}^T$, to compute the pseudoinverse $\mathbf{\Psi}^{\dagger}=\mathbf{\Psi}^T(\mathbf{\Psi}\mathbf{\Psi}^T)^{-1} = \mathbf{\Psi}^T(\hat{{\bf U}}_{\hat{r}}\hat{\mathbf{\Xi}}_{\hat{r}}\hat{{\bf U}}_{\hat{r}}^{\dagger})^{-1}=\mathbf{\Psi}^T\hat{{\bf U}}_{\hat{r}}\hat{\mathbf{\Xi}}_{\hat{r}}^{-1}\hat{{\bf U}}_{\hat{r}}^T$. Choose the truncation value $\hat{r}$.  
	\item Perform a second E-ED in the Hermitian space, ${\bf X}_1{\bf X}_1^T={\bf U}_{r}\mathbf{\Xi}_{r}{\bf U}_{r}^T$, to derive the POD modes (${\bf U}_{r}$) for reduced order projection. Choose the truncation value $r$ such that $\hat{r} > r$. 
	\item Compute the reduced order dynamic and input matrices: $\tilde{{\bf A}} = {\bf U}_{r}^T{\bf X}_2\mathbf{\Psi}\hat{{\bf U}}_{\hat{r}}\hat{\mathbf{\Xi}}_{\hat{r}}^{-1}\hat{{\bf U}}_{\hat{r},1}^T{\bf U}_{r}$ and $\tilde{{\bf B}} = {\bf U}_{r}^T{\bf X}_2\mathbf{\Psi}\hat{{\bf U}}_{\hat{r}}\hat{\mathbf{\Xi}}_{\hat{r}}^{-1}\hat{{\bf U}}_{\hat{r},2}^T$.
	\item Compute the DMD modes as	$\mathbf{\Phi}={\bf U}_{r}{\bf W}$, where ${\bf W}$ are the eigenvectors of $\tilde{{\bf A}}$ such that $\tilde{{\bf A}}{\bf W}={\bf W}\mathbf{\Lambda}$.
\end{enumerate}

\section{Model Development}\label{s:MD}

We use the TIE-GCM to perform the simulations for obtaining the snapshots. TIE-GCM is commonly used for modeling of the Ionosphere-Thermosphere environment. TIE-GCM is a comprehensive, first-principles based, three-dimensional, time-dependent, numerical simulation model of the Earth's upper atmosphere, including the thermosphere. It uses a finite differencing scheme to obtain a self-consistent solution for the three-dimensional momentum, energy, and continuity equations for neutral and ion constituents. The model can simulate on low and high resolution grid parameters. The model uses spherical geographic coordinates: latitude from -87.5 to 87.5 degrees (low resolution) and -88.75 to 88.75 degrees (high resolution), longitude from -180 to 180 degrees, with the vertical direction using a log-pressure coordinate system. The low and high resolution grids make increments of 5 and 2.5 degrees, respectively, in latitude and longitude \cite{TIE-GCM}. 

In this work, we run the model with standard inputs. The Heelis model of convection electric fields \cite{Heelis} is used, driven by the geomagnetic index Kp. Absorption of and ionization from solar ultraviolet is parameterized, driven by proxy in the form of the radio flux measured at a wavelength of 10.7 cm (F10.7). At the lower boundary of the model around 97 km, migrating diurnal and semidiurnal tidal fields are specified by the Global Scale Wave Model (GSWM) \cite{Hagan}, with eddy diffusivity specified in accordance with \cite{Qian}. 

We define the spatial decomposition of the thermosphere using a coarser resolution of the TIE-GCM grid. We reduce the grid to 24, 20, and 16 partitions in the geographic longitude, latitude and altitude dimensions, respectively in order to keep the problem computationally tractable. In addition, for this proof-of-concept, we use a little more than 105,000 snapshots with 1 hr resolution in time. Since TIE-GCM uses the log-pressure coordinate for the vertical dimension, the geometric height of the upper boundary can vary from $\sim$450-700 depending on solar activity. For the current work, we set the range of altitude at 100-450 km. Extending the model to higher altitudes will be a subject for future work. We convert from a log-pressure to a geometric height grid. We use density in the log scale for model development since its variance is much more uniform \cite{Emmert and Picone}. We then convert the azimuthal variable from longitude to local time since the local time variations are much larger than longitudinal variations.

Previous experience \cite{Mehta_POD} suggests that in order for the developed ROM to be applicable for all input conditions, the snapshots need to include simulation output covering the full input domain. However, as a first step, it is important to get a feel for the effectiveness of HS-DMDc for the problem of interest under an observed input scenario. Therefore, we first apply the HS-DMDc algorithm to TIE-GCM simulations performed with observed inputs for the years 2002 and 2008, representing high and low solar activity conditions, respectively. 

\subsection{Method Validation}\label{s:MeV}

\begin{figure}[h]
	\centering
	\includegraphics[width=\textwidth]{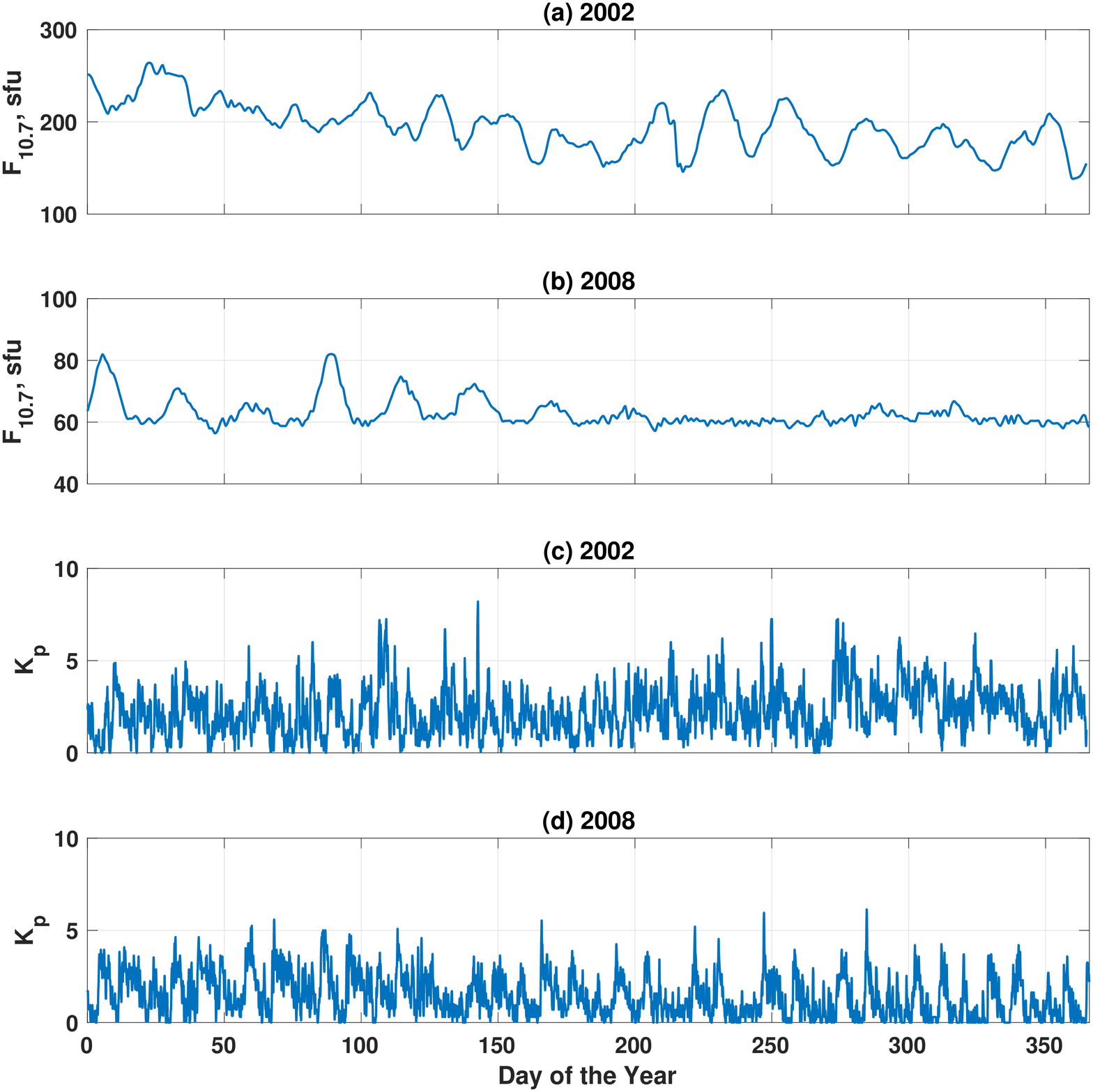}
	\caption{$F_{10.7}$ and $K_p$ inputs for years 2002 and 2008}
	\label{f:2002_2008_Inputs}
\end{figure}

Figure~\ref{f:2002_2008_Inputs} shows the variation of $F_{10.7}$ and $K_p$ for years 2002 and 2008. The 27-day period is clearly visible in $F_{10.7}$ for both years while the $K_p$ variation seems purely stochastic. The year 2002 was slightly more active geomagnetically, having some of the largest geomagnetic storms of the solar cycle, with an average $K_p$ close to 3 whereas the average for 2008 is closer to unity. Also, the $K_p$ values reach extremely high values (> 6) during 2002, albeit, very rarely. 

\begin{figure}[htb!]
	\centering
	\includegraphics[width=\textwidth]{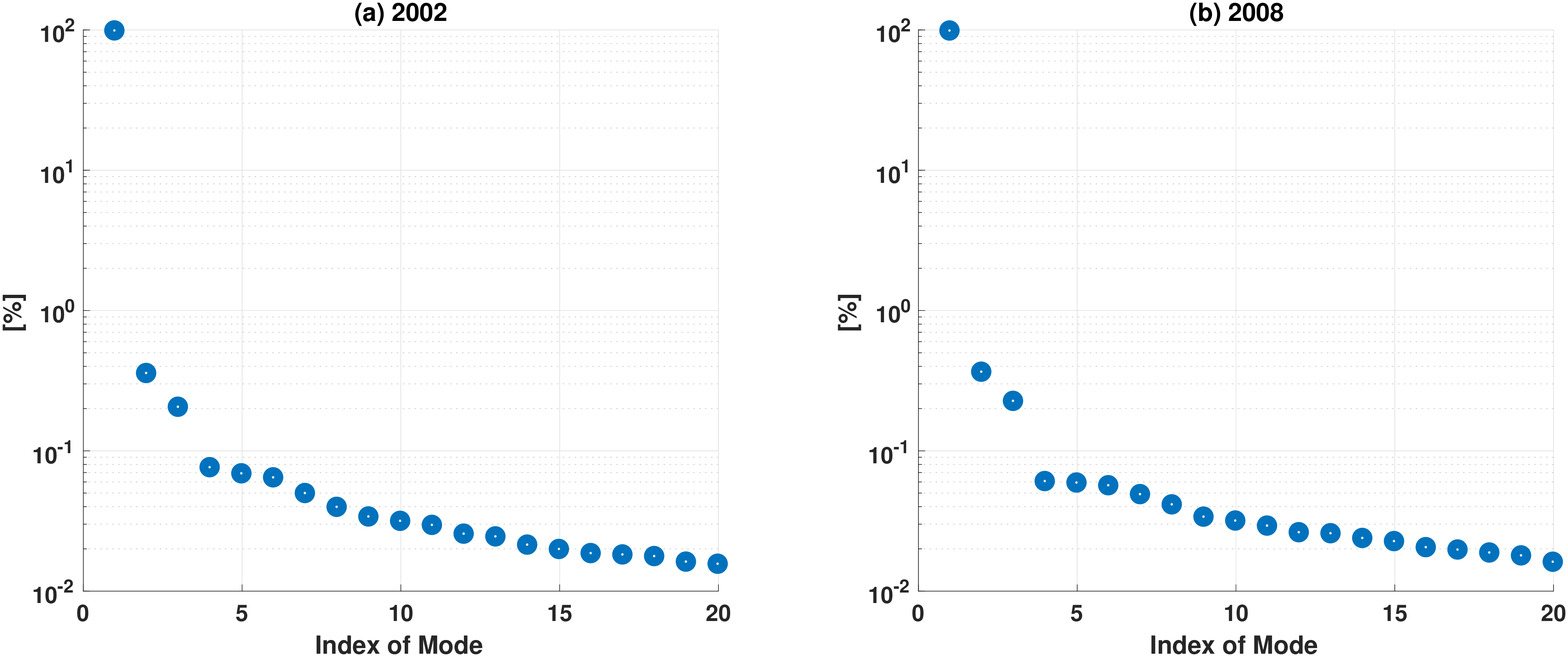}
	\caption{The contriubtion of the first 20 POD modes to the total energy for 2002 and 2008. The first 20 modes capture more than 99\% of the total energy.}
	\label{f:EV_2002_2008}
\end{figure}

\begin{figure}[htb!]
	\centering
	\includegraphics[width=\textwidth]{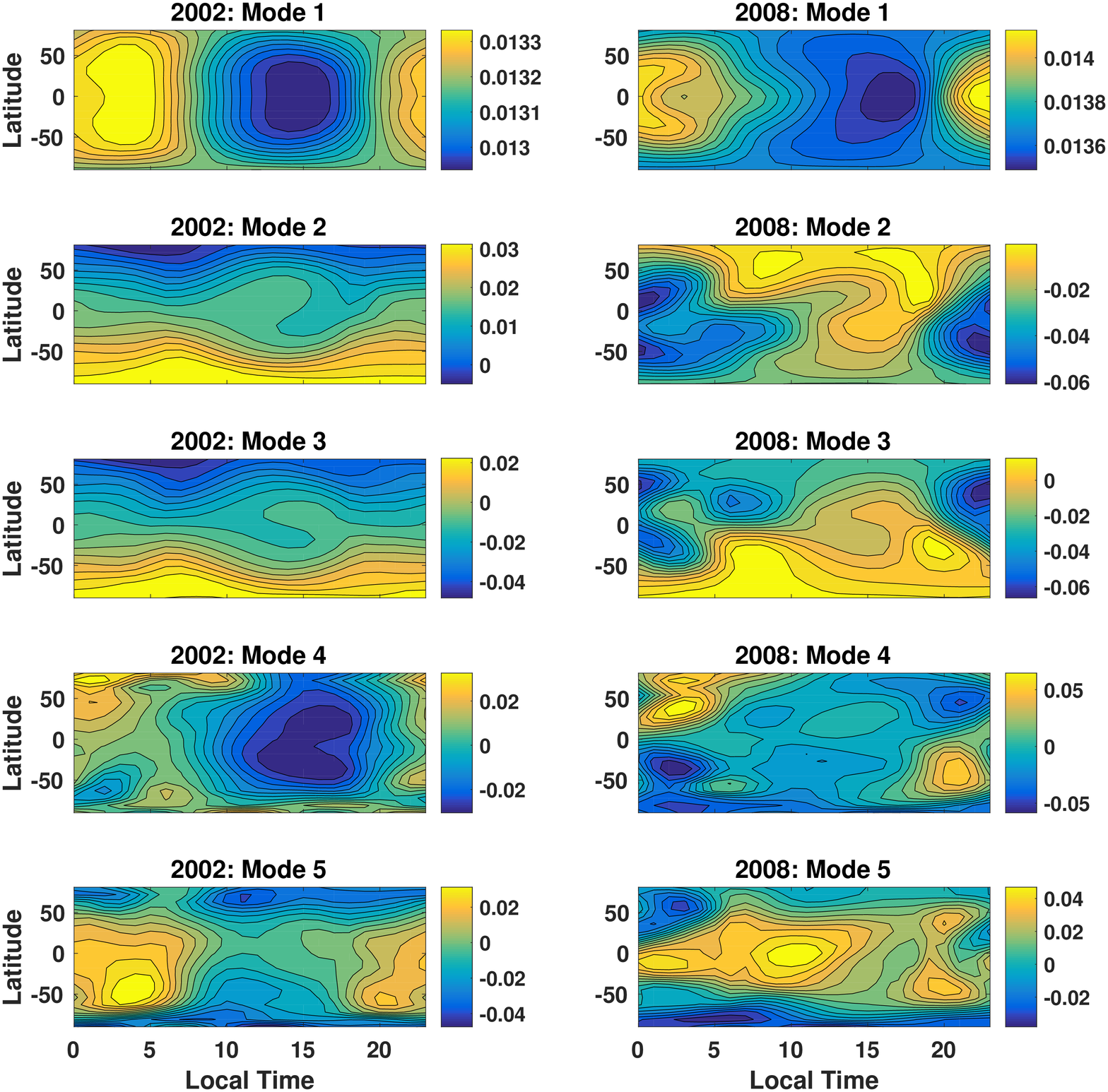}
	\caption{First 5 POD modes for 2002 and 2008 derived using HS-DMDc at 450 km altitude.}
	\label{f:Modes_2002_2008}
\end{figure}

We apply the HS-DMDc algorithm to snapshots collected every hour for both years. The choice of truncation value $r$ is subjectively constrained to $\hat{r}>r$. While the choice of $\hat{r}$ needs to ensure that all important modal excitations by the inputs are captured, the choice of $r$ is a trade-off between accuracy and model parameter estimation for data assimilation, which will be a subject of future work. We use four inputs ($p$), the solar flux proxy $F_{10.7}$, the geomagnetic index $K_p$;  time of the day in universal time $UT$, and day of the year $DOY$. Figure~\ref{f:EV_2002_2008} shows the contribution of the first 20 POD modes to the total energy upon which the full state dynamics matrix is projected for order reduction, while Figure~\ref{f:Modes_2002_2008} shows the first 5 POD modes sampled at 450 km altitude for 2002 and 2008. Mode 1 which includes a strong component of the mean, accounts for about 98\% of the total energy. Qualitative analysis of the modes is beyond the scope of the current work, but quick observations can be made based on the comparison of the modes for 2002 and 2008. Temperature through $F_{10.7}$ dominates the variations; the modes for 2002 in Figure~\ref{f:Modes_2002_2008} are similar to those derived by \cite{EOF1} using CHAMP and GRACE derived density observations. Modes 2 and 3 for 2002 represent annual and semi-annual variations; however, mode 2 is also influenced solar activity ($F_{10.7}$) that drives the temperature variations. It is important to note that the dominant modes are different for 2002 and 2008 due to the difference in solar activity levels. Mode 2 for 2008 also seems solar activity driven with a weak diurnal component. Mode 3 for 2008 seems to represent annual variation but also with an influence of solar activity. 

\begin{figure}[htb!]
	\centering
	\includegraphics[width=\textwidth]{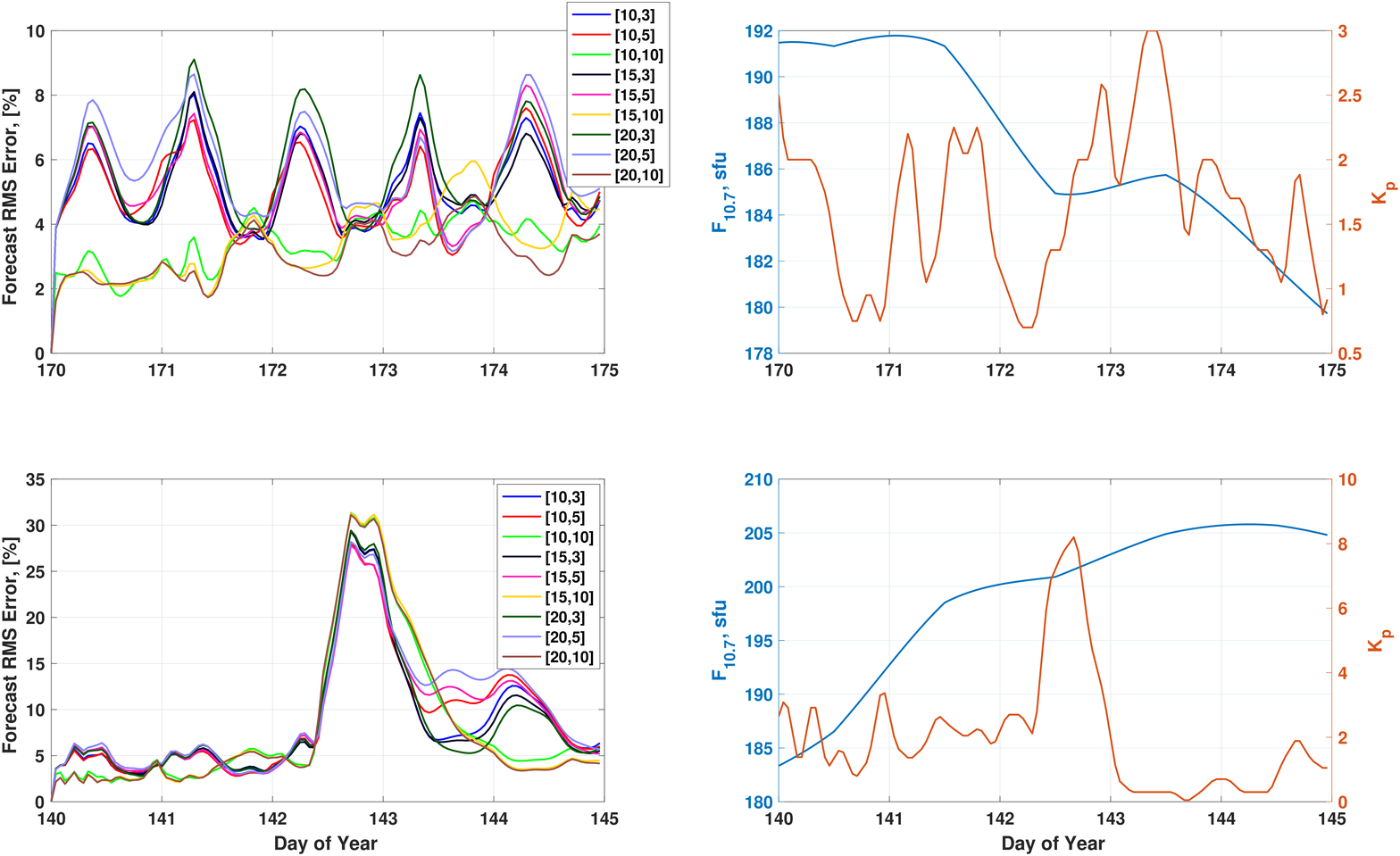}
	\caption{Forecast error in reproducing the 2002 simulation snapshots using ROM developed with the same 2002 snapshots for different combinations of [$\hat{r}, r$].}
	\label{f:2002_170_140_FE}
\end{figure}

\begin{figure}[htb!]
	\centering
	\includegraphics[width=\textwidth]{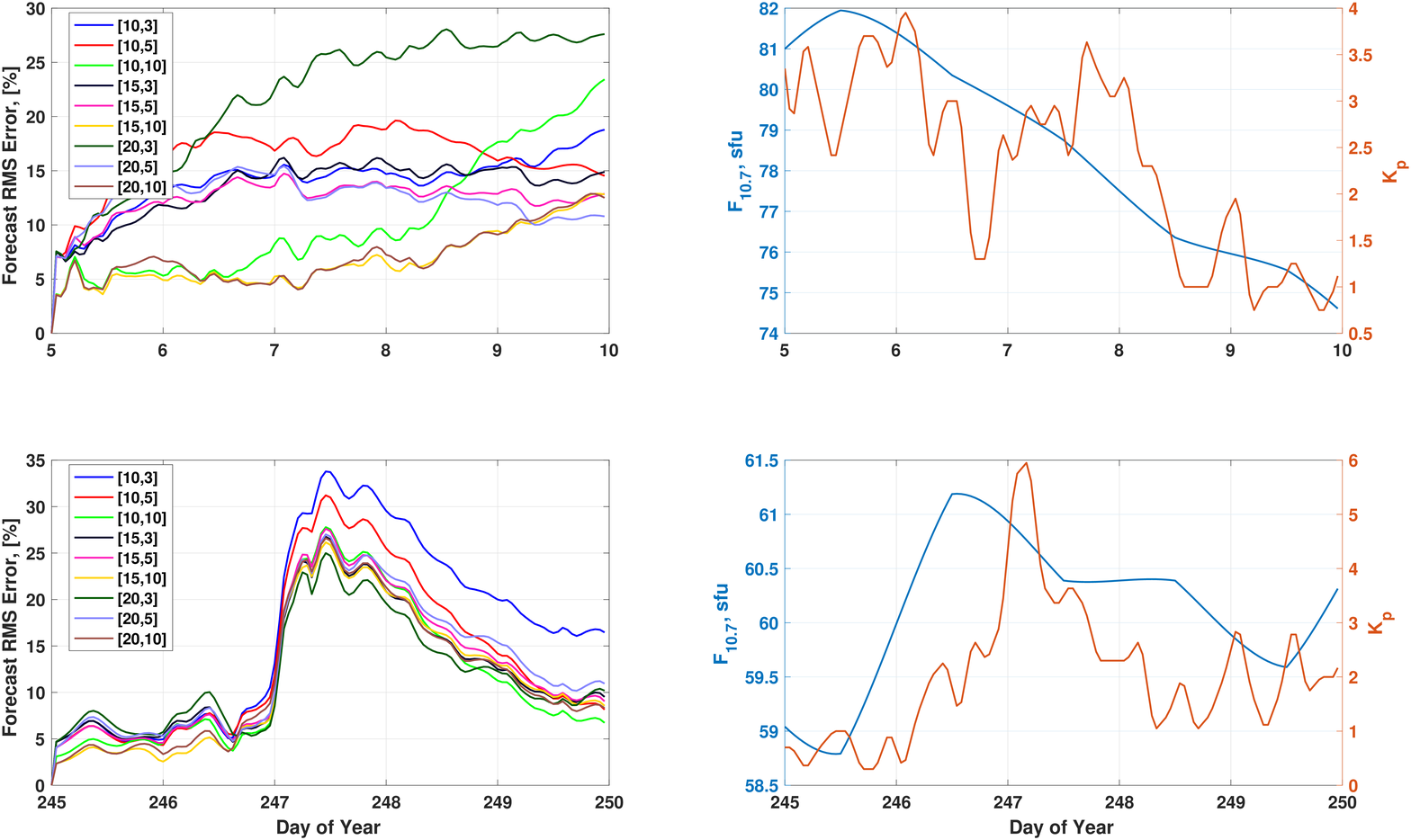}
	\caption{Forecast error in reproducing the 2008 simulation snapshots using ROM developed with the same 2008 snapshots for different combinations of [$\hat{r}, r$].}
	\label{f:2008_5_245_FE}
\end{figure}

For both years, the first 20 modes capture 99\% of the total energy. Next, we perform a sensitivity analysis to understand the impact of truncation values $\hat{r}$ and $r$. We vary $r$ to take values of 3, 5, and 10. Because $\hat{r}>r$, we vary $\hat{r}$ to take values of 10, 15, and 20. Figures~\ref{f:2002_170_140_FE} and \ref{f:2008_5_245_FE} show the results of sensitivity analysis for 2002 and 2008, respectively. The forecast error represents the state propagation error, given initial condition and inputs. The forecast error is the root mean square percentage error on the 3-dimensional spatial grid. The top panels in the figures cover inputs close to the mean for the year. The bottom panels cover inputs at the extremes in either $F_{10.7}$, $K_p$, or both.

As expected, state propagation using the locally linear approximation of the dynamic matrix in time causes the solution to depart from the true solution; however, the forecast error after 1 day for each case is close to or below 10\%. The forecast errors seem to band based on the number of modes used for order reduction, $r$, with very small sensitivity to $\hat{r}$. A value of $\hat{r}=$ 20 seems to capture all of the important excitations by the inputs with forecast errors for $\hat{r}=$ 15 and 20 being nearly indistinguishable. The rest of the paper will present results using $\hat{r}=$ 20 and $r=$ 10, unless otherwise specified. The forecast errors rise faster for decreasing values of $r$. Figures~\ref{f:2002_170_140_FE} and \ref{f:2008_5_245_FE} also show that the errors rise and fall sharply around extreme values of $K_p$, possibly due to the snapshots not covering the high $K_p$ enough times and/or the model becoming highly non-linear. Within data, the current approach shows comparable performance for the case with low geomagnetic activity as the POD-based approach discussed in \cite{Mehta_POD}. However, the current approach is much simpler and provides a natural extension for forecasting. The model performance at really high values of $K_p$ is discussed in a later section as a limitation of the model.

\begin{figure}[htb!]
	\centering
	\includegraphics[width=\textwidth]{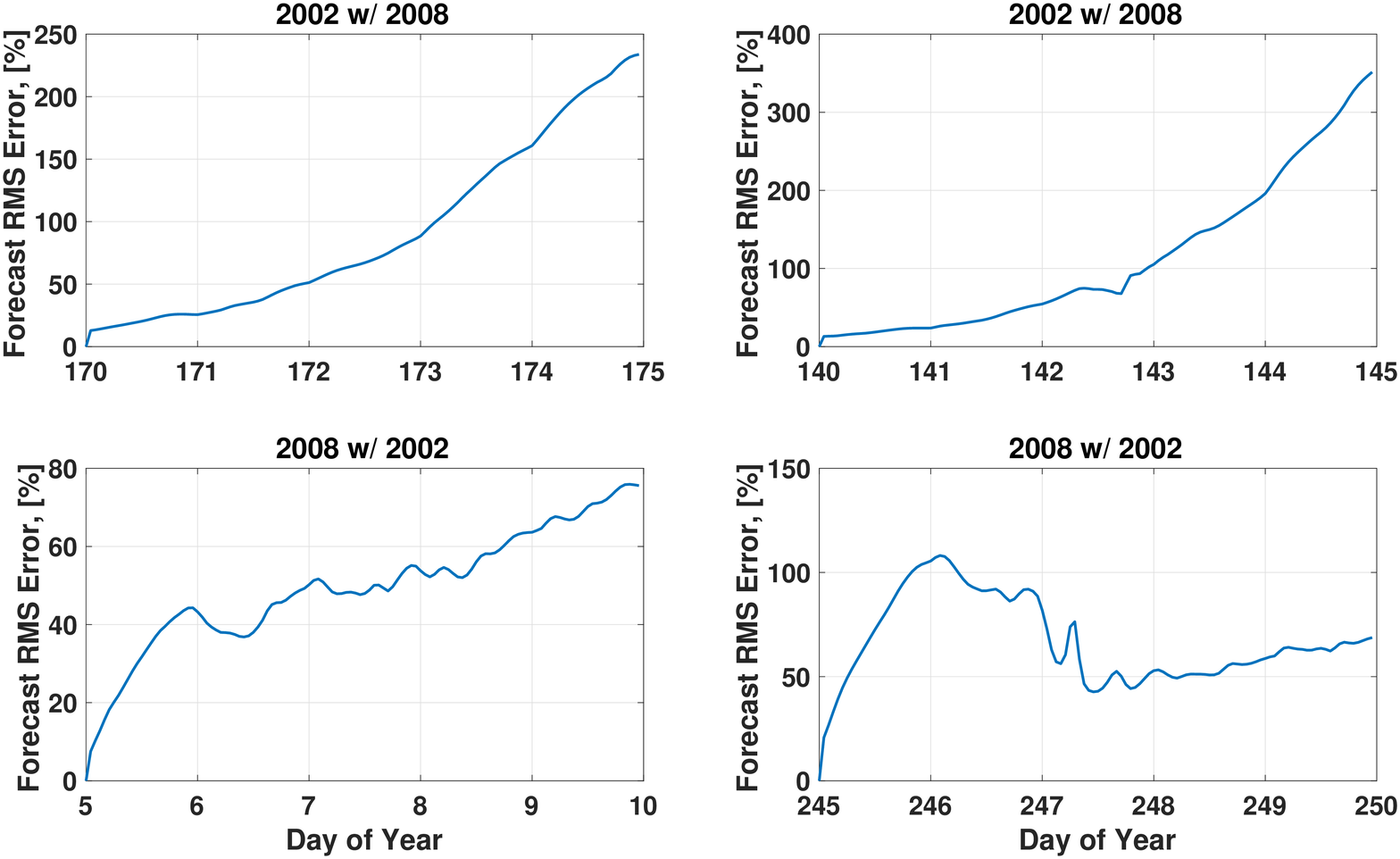}
	\caption{Cross Prediction forecast error: 2002 snapshot reproduction with model generated using 2008 snapshots and vice-a-versa.}
	\label{f:Cross_Prediction}
\end{figure}

To assess the ability of the technique to specify density outside the range of conditions under which it was tuned, we attempt to reproduce the 2008 (low solar activity) snapshots using the 2002 model (high solar activity) and vice-a-versa. Figure~\ref{f:Cross_Prediction} shows the results of this cross prediction. As observed, the forecast errors are much larger than the errors in Figures~\ref{f:2002_170_140_FE} and \ref{f:2008_5_245_FE} and rises in general propagating away from the initial condition. Further investigation into the cause behind the large errors affirms previous knowledge that the developed models are only applicable for conditions captured by the snapshots. Figure~\ref{f:Cross_Prediction} shows errors that converge (or close to convergence) to large steady state values that represent the inability to match the variations caused by solar activity (scaling by $F_{10.7}$) because the mean component in mode 1 only applies to the solar activity levels covered with the snapshots. In other words, the model captures most of the dynamics but is biased in the absolute scale. An exception is observed in Figure~\ref{f:Cross_Prediction}(d) where the error rises quickly due to the bias in the absolute density during high and low levels of solar activity, but falls due to the geomagnetic storm significantly increasing the density taking it close to 2002 levels. The error again rises after the storm has passed. 

\subsection{Universal Model}\label{s:UM}

As mentioned previously, in order for the developed model to be universally applicable, the snapshots need to cover the full range of input conditions with the snapshots. Covering the entire range of $F_{10.7}$ with true variations along with the condition for the snapshot matrices to be dynamically continuous requires TIE-GCM simulations for more than 10 years (a full solar cycle), which in itself can be computationally expensive. Therefore, we first develop two models using input sampling strategies needed to ensure that the snapshots cover all possible input conditions with only one year of TIE-GCM simulations. We then develop a third model that uses 12 years of TIE-GCM simulations covering the last solar cycle from 1997 to 2009. 

The standard methods used for dynamic system identification are not expected to work for the problem at hand since \textit{time} is an input, e.g. annual and semiannual variations. Therefore, one straightforward strategy to ensure adequate coverage is to use oscillatory functions (e.g., sine or cosine) for both $F_{10.7}$ and $K_p$. We first attempted a sine function for $F_{10.7}$ and $K_p$ with periods of 27 and 4 days respectively. However, inspection of the POD modes showed that an oscillatory $K_p$ resulted in artificial modes with periods of 4 days resulting in large model (defined as the error at current time, $k$, given initial condition and control at previous time, $k-1$) and forecast errors. This sits well with the understanding that $F_{10.7}$, for the most part, scales the atmosphere (contraction and expansion) without significantly modifying the global distribution, whereas $K_p$ very sensitively impacts the global distribution captured by the spatial modes. 


With that knowledge, we perform two additional runs of TIE-GCM to collect snapshots both using oscillatory $F_{10.7}$ and Sim1: $K_p$ sampled from observed distributions,  and Sim2: $K_p$ sampled from normal distributions, one each for high and low solar activity. Recognizing that persistence is a strong indicator of geomagnetic activity, at least on relatively short timescales (i.e., 3 hours or less), we developed a strategy to construct a random time series of $K_p$ that preserves this quality. For our $K_p$ sampling, we analyzed historical 3-hourly $K_p$ values dating back to 1947 obtained from NOAA's National Center for Environmental Information (NCEI). These values were then separated into low, medium, and elevated solar activity levels based on the corresponding daily $F_{10.7}$ values with following respective binning thresholds:  $F_{10.7} < 120$ sfu, $120 \leq F_{10.7} < 180$ sfu, and $ F_{10.7} \geq 180$ sfu. Within each of these three bins, a conditional probability distribution was constructed for $K_p(k)$ at the current time $t$, based on the previous 3-hourly value, $K_p(k-1)$. Given the 28 discrete values that $K_p$ can take, this method requires construction of 84 separate histograms. A time series can then be created by randomly drawing from these histograms, knowing the current $F_{10.7}$ and seeding with an initial $K_p$ value. This method produces a timeseries in which both the conditional and overall distribution of $K_p$ probabilities resembles those of the historical $K_p$ indices.

\begin{figure}[htb!]
	\centering
	\includegraphics[width=\textwidth]{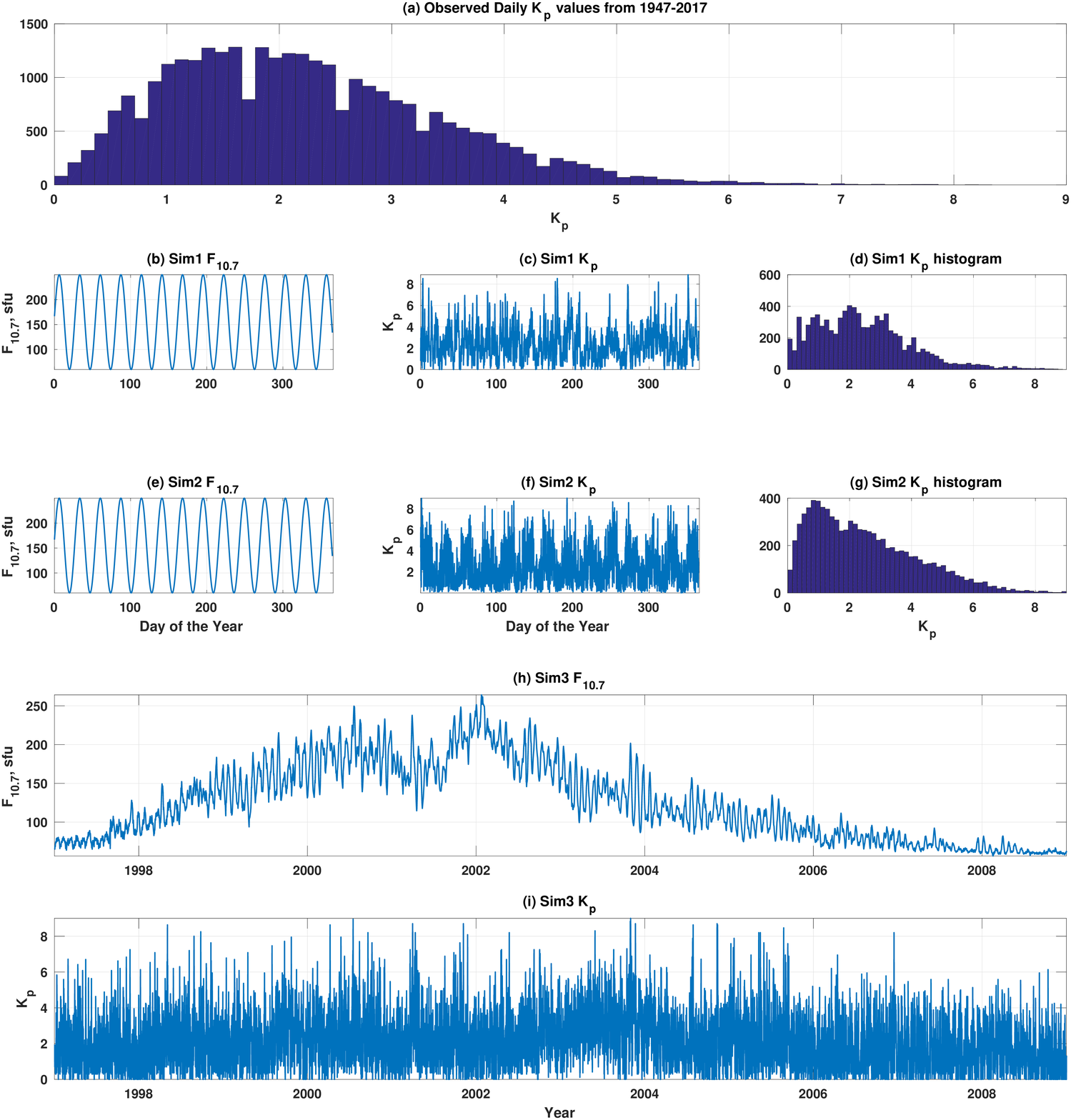}
	\caption{(a) Histogram of the observed daily $K_p$ values from 1947-2017. (b,e) simulated oscillatory $F_{10.7}$ input for Sim1 and Sim2. (c,f) simulated $K_p$ input for Sim1 and Sim2. (d,g) histograms of the simulated $K_p$ inputs for Sim1 and Sim2. (h) and (i) shows the $F_{10.7}$ and $K_p$ variation from 1997-2009 used in Sim3, respectively. }
	\label{f:Sim123_Inputs}
\end{figure}

Sim2 uses two normal distributions: $[\mu,\sigma]=[0,2.5]$ for $F_{10.7} < 120$ sfu and $[\mu,\sigma]=[3,3]$ for $F_{10.7} > 120$ sfu, where $\mu$ is the mean and $\sigma$ is the standard deviation. We apply several operators to the distributed samples to make them more physically realistic. We convert the sampled negative values to positive using the absolute operator. In addition, samples above nine are converted to a value equal to nine. We then subtract half of the maximum value from samples with values above 4.5 when $F_{10.7}<120$. This is because active geomagnetic conditions are typically observed for an active Sun. Sim3 uses the observed $F_{10.7}$ and $K_p$ for the period from 1997 to 2009. Figure~\ref{f:Sim123_Inputs} shows the distribution of observed $K_p$  since 1947, simulated inputs for cases Sim1 and Sim2, and the variation of the observed $F_{10.7}$ and $K_p$ from 1997 to 2009 used in Sim3. As observed, the $K_p$ distribution for Sim2 pushes the tail further to the right with more samples in the 4-5 range. This allows to expand the region where the model is applicable as will be discussed in section \ref{s:ML} on model limitations. While the peak of the observed $K_p$ distribution lies close to a value of 1.5, the peaks for Sim1 and Sim2 lie close to values of 2 and 1, respectively. 

\begin{figure}[htb!]
	\centering
	\includegraphics[width=\textwidth]{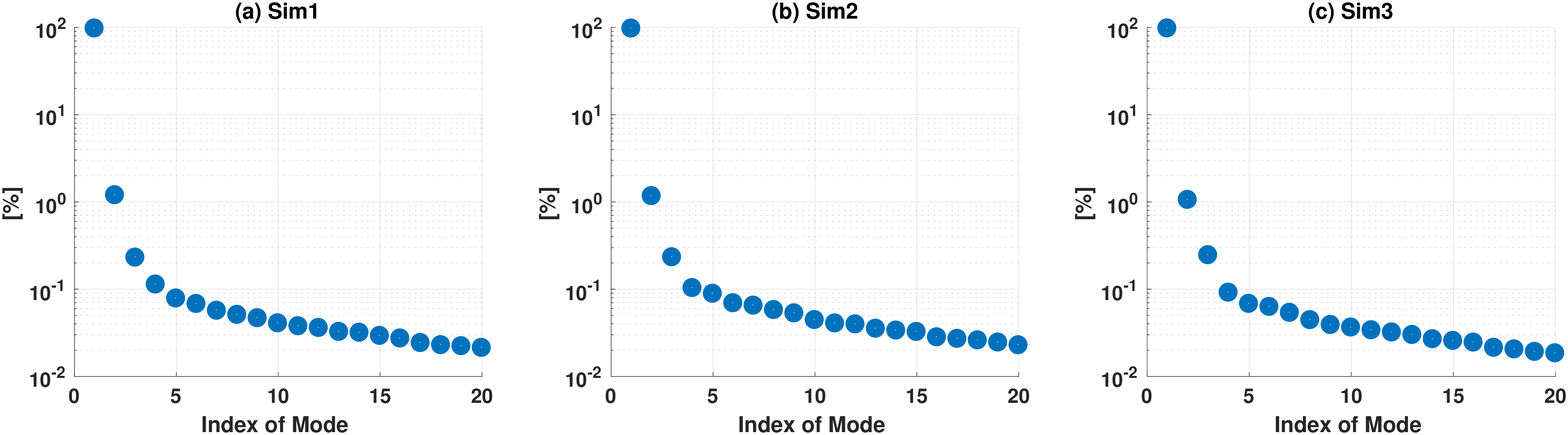}
	\caption{The contriubtion of the first 20 POD modes to the total energy for Sim1, Sim2, and Sim3. The reduced order dynamic matrix $\tilde{{\bf A}}$ is obtained by projecting the full order dynamic matrix ${\bf A}$ onto the POD modes. The first 20 modes capture more than 99\% of the total energy.}
	\label{f:Sim123_EVs}
\end{figure}

\begin{figure}[htb!]
	\centering
	\includegraphics[width=\textwidth]{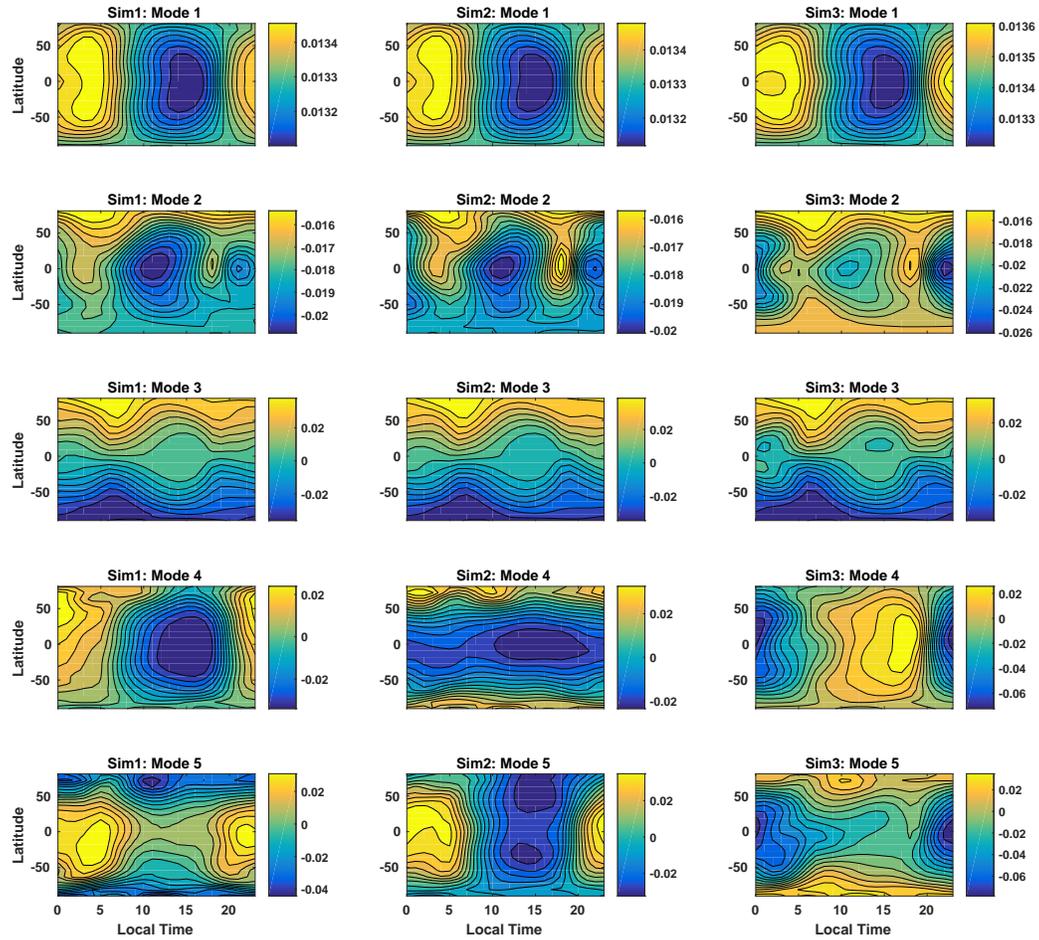}
	\caption{First 5 POD modes for Sim1, Sim2, and Sim3 derived using HS-DMDc at 450 km altitude.}
	\label{f:Sim123_POD_Modes}
\end{figure}

\begin{figure}[htb!]
	\centering
	\includegraphics[width=\textwidth]{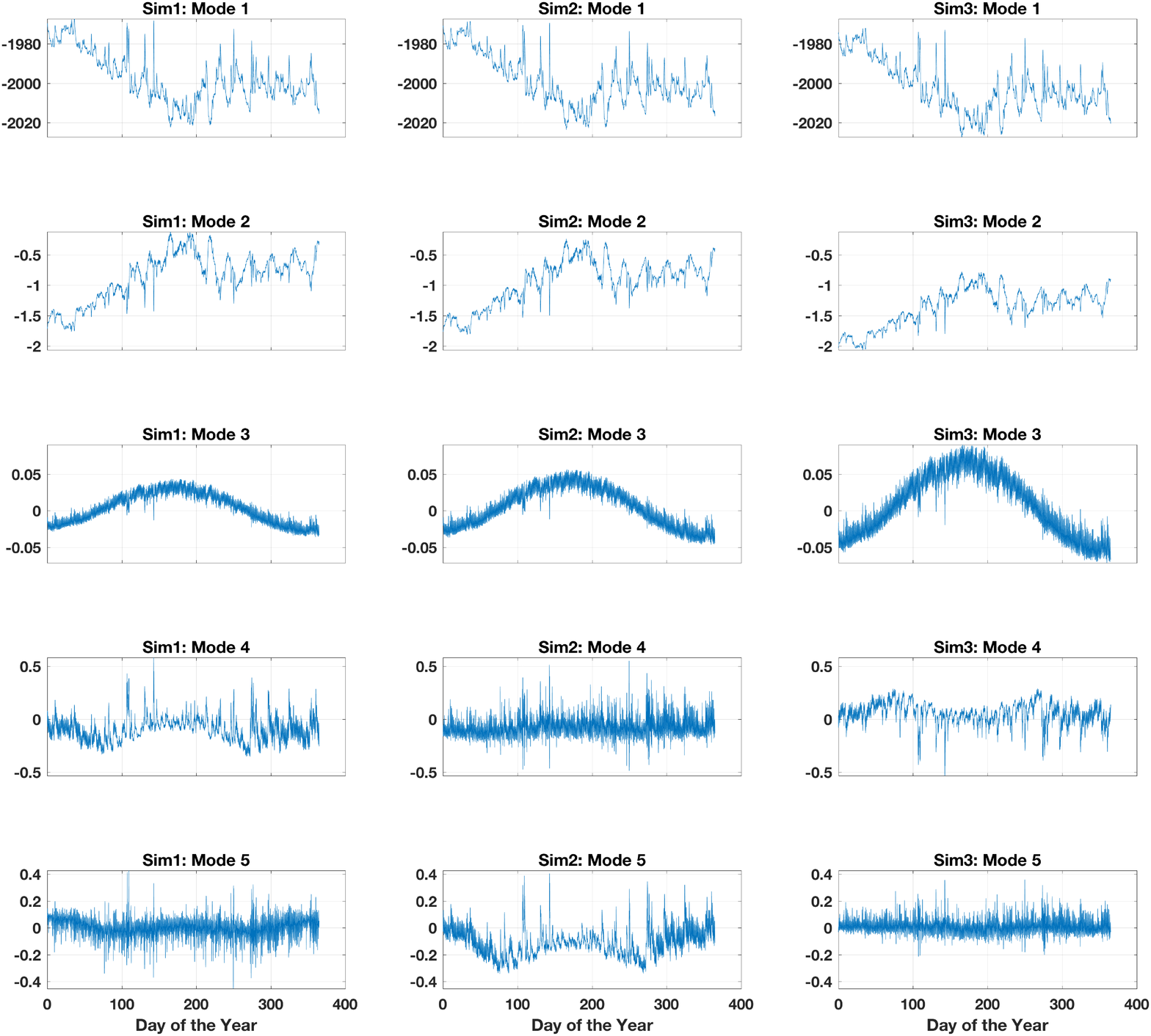}
	\caption{Coefficients for first 5 POD modes for Sim1, Sim2, and Sim3. Sim3 coefficients are for the year 2002.}
	\label{f:Sim123_POD_Coefficients}
\end{figure}

Figure \ref{f:Sim123_EVs} shows the normalized POD eigenvalues that represent the contribution to the total energy for the three cases. For all the three cases, the first mode captures close to 97\% with the first 3 modes capturing more than 98\% of the total energy. Figure~\ref{f:Sim123_POD_Modes} shows the first 5 POD modes for the three cases. As previously mentioned, qualitative analysis of the dynamics is beyond the scope of the current work; therefore, we will only make some quick observations. Mode 1 again contains a strong mean component with the first 3 modes being almost identical. Figure~\ref{f:Sim123_POD_Coefficients} shows the coefficients corresponding to the first 5 modes. Sim3 POD coefficients shown are for the year 2002. Modes 1 and 2 have a strong correlation (positive and negative, respectively) with $F_{10.7}$ and the same 27-day period representing the variations caused by solar activity. The third mode represents annual variation. Mode 4 for Sim1 and Sim3 and mode 5 for Sim3 represent semi-annual variations. Mode 5 for Sim1 and Sim2 and mode 4 for Sim3 also contain a very weak semi-annual trend.

\subsection{Universal Model Validation}\label{s:UMV}

\begin{figure}[htb!]
	\centering
	\includegraphics[width=\textwidth]{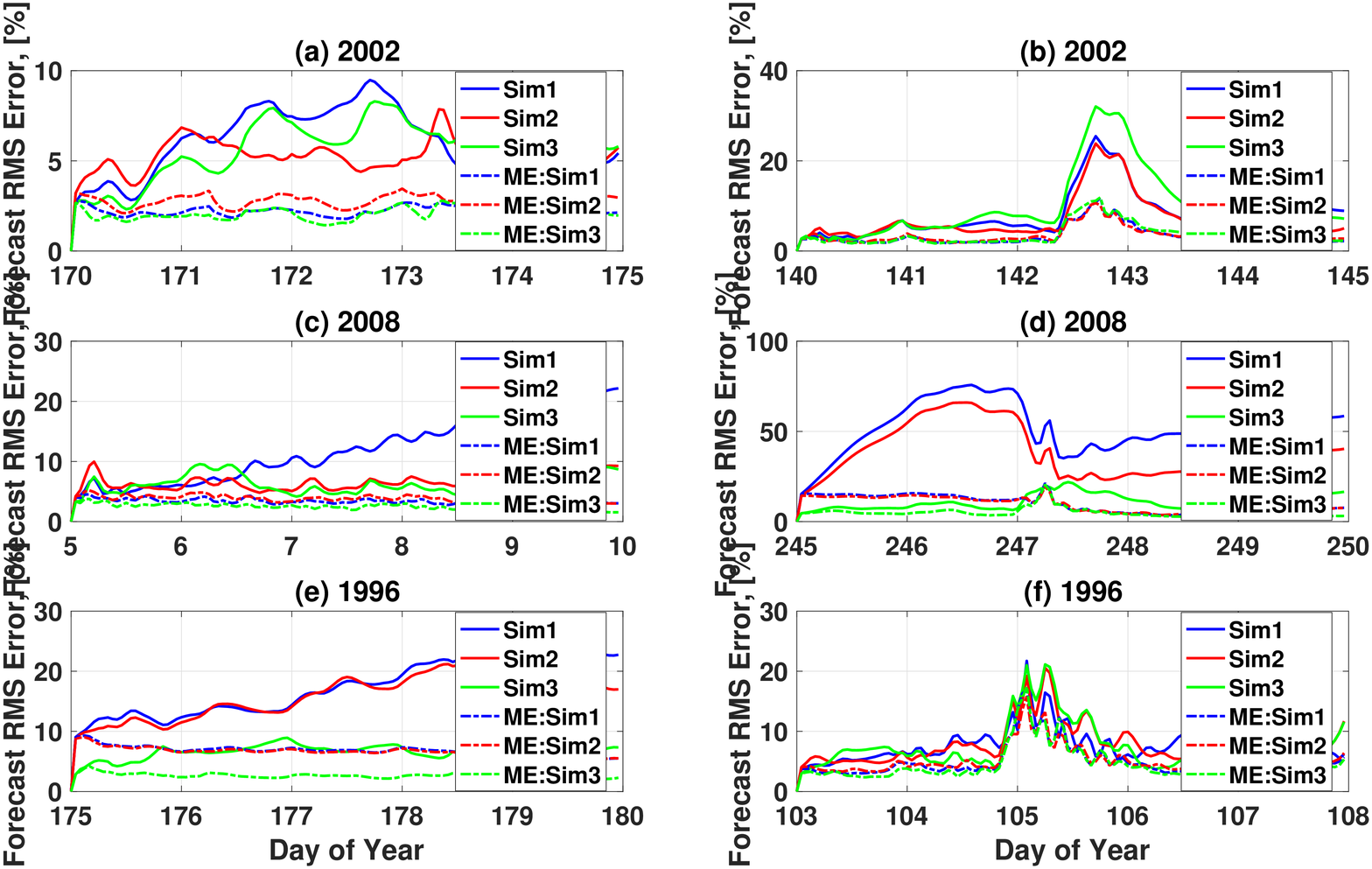}
	\caption{Validation of Sim1 and Sim2 models. ME: Model Error is defined as the error at current time $k$ given initial condition and control at previous time $k-1$ }
	\label{f:Sim123_Validation}
\end{figure}

We validate the three (Sim1, Sim2, and Sim3) models with the real case simulations performed for 2002 and 2008. Because TIE-GCM output for 2002 and 2008 is incorporated through the snapshots into the Sim3 model, we perform additional TIE-GCM simulations for the year 1996 to conduct a validation of the models with an independent dataset. Figure~\ref{f:Sim123_Validation} shows the forecast error using Sim1, Sim2, and Sim3 models for a couple different initial conditions in 2002, 2008, and 1996, respectively. We use the initial conditions for 2002 and 2008 and choose two initial conditions for 1996 that represent mean inputs and extreme conditions as previously described in section~\ref{s:MeV}. As observed, the Sim1 and Sim2 ROMs perform well in keeping the forecast error close to or below 10\% after 24 hours and close to or below 15\% after 48 hours in general. An exception is shown in Figure~\ref{f:Sim123_Validation}d, where much like in Figure~\ref{f:Cross_Prediction}d, the forecast error is initially very large and falls with the onset of the storm and begins to rise again after the storm due to the dynamic propagation. The large errors in the first 24-48 hours are because the $F_{10.7}$ value, and hence the density, for the first 48 hours is outside or on the boundary of the input range incorporated in the Sim1 and Sim2 snapshots. The error falls with the onset of the storm because it increases the density and brings it in the range of the values incorporated into Sim1 and Sim2. It is important to note that the Sim1 and Sim2 absolute errors in Figure~\ref{f:Sim123_Validation}d are much smaller than the errors in Figure~\ref{f:Cross_Prediction}d because of the sampling strategies used. The errors can be reduced by expanding the range of the inputs. 

Sim3 consistently outperforms Sim1 and Sim2 in every scenario including the validation using independent datasets of 1996. Sim3 ROM performs very well in keeping the forecast error close to or below 5\% after 24 hours and close to or below 10\% after 48 hours in general. Sim3 also performs well in Figure~\ref{f:Sim123_Validation}d because it incorporates the 2008 TIE-GCM output into the model through the snapshots. The model errors for Sim3 are also either better or as good as for Sim1 and Sim2. For all the three models, the model and forecast errors rise sharply during a storm event. The models can be improved either by incorporating more storm-time snapshots into the development or through data assimilation, which will be the subject of future work. The storm time performance is a limitation for the current version of the model.

Both the forecast and model errors for all the three model can be further reduced by increasing the truncation order $r$. As alluded to previously, the choice of $r$ is a trade-off between accuracy and the practicality of estimating model parameters (modal coefficients) for data assimilation. By definition, the idea behind modal decomposition is to capture a significant portion of the energy/variance with a small number of modes, ideally less than 5. Even for application to reduced order modeling of the upper atmosphere, close to 98\% of the total energy is captured by the first 5 modes. When assimilating measurements, using a large number of modes can lead to overfitting. However, certain dynamics that are important but do not dominate the energy/variance may not be captured in the first 5-10 modes. This argument and trade-off will be discussed in detail in future work on the development of the framework for data assimilation for the ROM.

\subsection{Dynamic modes }\label{s:DM}

\begin{figure}[htb!]
	\centering
	\includegraphics[width=\textwidth]{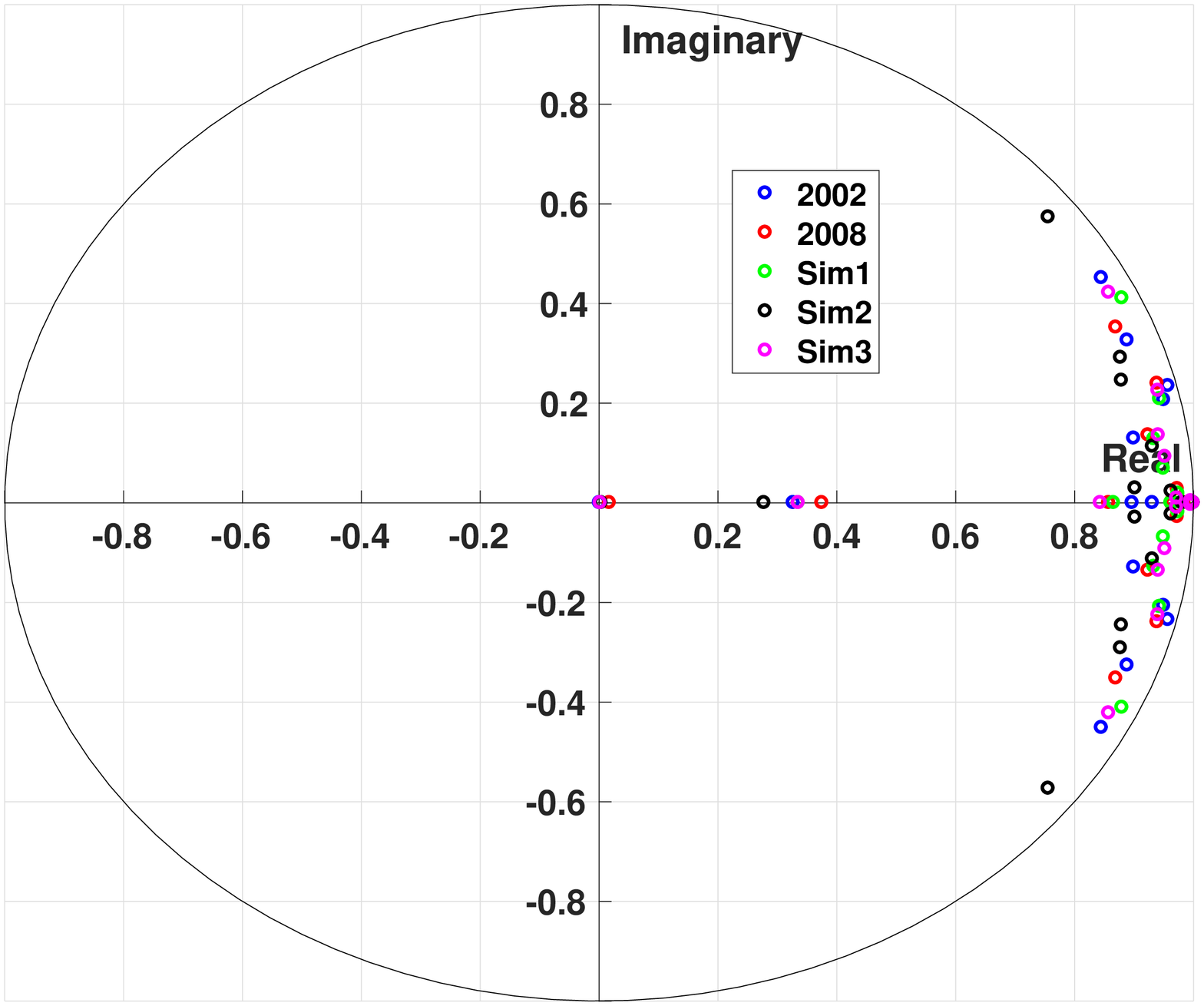}
	\caption{Discrete-time DMD eigenvalues of the reduced order dynamic matrix $\tilde{{\bf A}}$ for 2002, 2008, Sim1, Sim2, and Sim3.}
	\label{f:Sim123_DMD_EVs}
\end{figure}

Although the DMD modes (computed in step 5 of the HS-DMDc in section \ref{s:NA}) do not play a role in the development of the ROM, we compute the DMD modes because they provide insights into the dynamics embedded within the dynamic matrix ${\bf A}$. In addition, detailed qualitative analysis of the dynamics is beyond the scope of this work; however, the DMD eigenvalues provide a measure of thermosphere dynamics on a wide range of time scales. Figure~\ref{f:Sim123_DMD_EVs} shows the discrete-time eigenvalues of the reduced order dynamic matrix for the models corresponding to 2002, 2008, Sim1, Sim2, and Sim3. The eigenvalues are a mix of real and complex conjugate pairs. As observed, there is some overlap of the DMD eigenvalues, representing the common set of dynamics between the models. The eigenvalues most likely do not overlap exactly because of dynamics that differ as a function of the input conditions. The angle from the real axis for an eigenvalue represents the single frequency of the dynamic mode, whereas, the magnitude (distance from the origin) represents the decay rate for the given mode. As observed, most eigenvalues are clustered close to the unit circle; a magnitude greater than unity corresponds to an unstable or growing mode, a value of unity corresponds to modes with no growth or decay, while a value less than unity corresponds to a decaying mode. An eigenvalue on the real line corresponds to pure scaling with no oscillatory properties.

\begin{figure}[htb!]
	\centering
	\includegraphics[width=\textwidth]{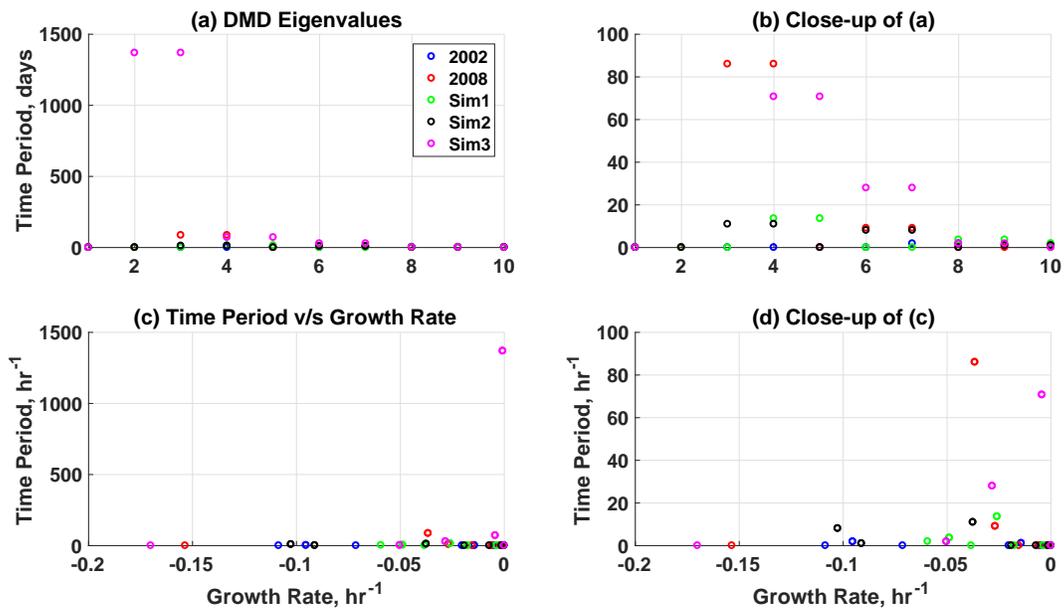}
	\caption{Continuous-time time period (1/frequency) and growth-rates for the DMD modes of the reduced order dynamic matrix $\tilde{{\bf A}}$ for 2002, 2008, Sim1, Sim2 and Sim3 ordered by the custom criteria described above. Time periods (Inf) corresponding to zero frequency are not set to zero.}
	\label{f:Sim123_DMD_TPs}
\end{figure}

\begin{figure}[htb!]
	\centering
	\includegraphics[width=\textwidth]{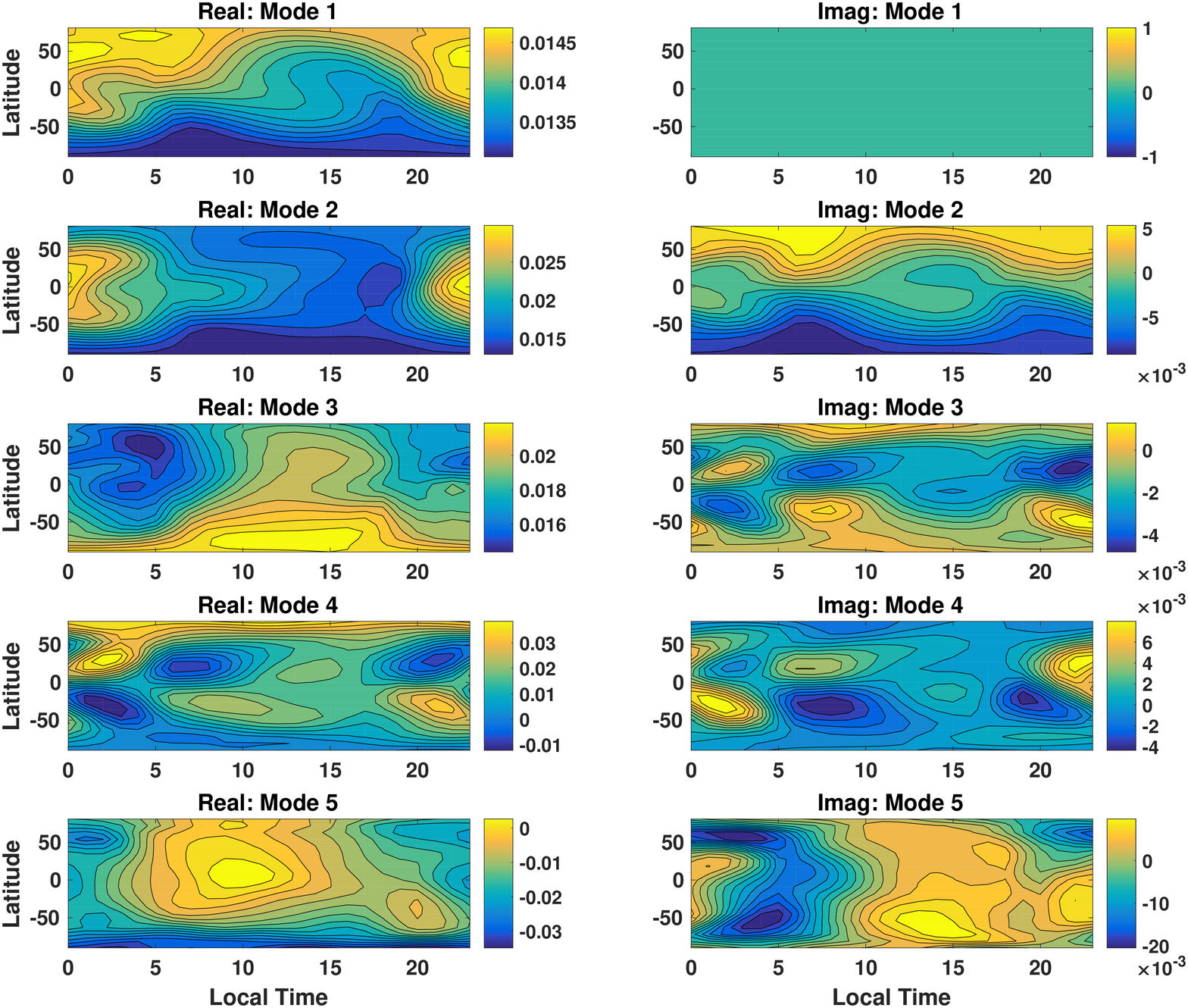}
	\caption{The real and imaginary parts for the DMD modes of the reduced order dynamic matrix $\tilde{{\bf A}}$ for Sim3 at 450 km altitude.}
	\label{f:Sim3_DMD_Modes}
\end{figure}

The discrete-time eigenvalues in Figure~\ref{f:Sim123_DMD_EVs} can be converted to continuous-time using the relation in Eq.~\ref{e:DS5}. Because the DMD eigenvalues and eigenvectors computed with the eigendecomposition are not ordered in any given order, we sort the non-zero (|$\Lambda$| > 1e-3) eigenvalues and eigenvectors using a custom criteria. We sort by total contribution of the DMD mode calculated by projection of the eigenvectors onto the first 10 POD modes scaled by the POD eigenvalues. Figure~\ref{f:Sim123_DMD_TPs} shows the continuous-time time period (1/frequency; value of Inf for time period corresponding to a zero frequency is set to 0) and growth rates of the dynamic modes ordered using the sorting described above. Figures~\ref{f:Sim3_DMD_Modes} show the first 5 dynamic modes for Sim3. To avoid redundancy, we show only one mode for each complex conjugate pair. Dynamic modes for the other models are not shown to save on space. Corresponding to the eigenvalues, the dynamic modes can be real or complex; the real modes or eigenvalues represent pure scaling with no dynamic or oscillatory properties. The real and imaginary parts combine to give the modes their dynamic properties. 

The first dynamic mode is real with a corresponding zero frequency and represents scaling of the atmosphere with $F_{10.7}$. Modes 2 represents a conjugate pair and corresponds to scaling and long term variations driven by $F_{10.7}$. The time period of close to 4 years seems to be a result of the variation of $F_{10.7}$ with a dip at close to 4 years (at the beginning of 2001). We expect that the conjugate pair would manifest as independent real modes in the absence of the dip in $F_{10.7}$ (the first 5 modes for 2002 are all real). Mode 3 represents a conjugate pair combining semi-annual variations and higher order harmonics of the daily period. The combination has a time period of close to 70 days. Mode 4 represents a conjugate pair that seems to correspond to the 27-day variations in $F_{10.7}$. The next three modes, including Mode 5 in Figure~\ref{f:Sim3_DMD_Modes}, seem to represent short term variations. All the dynamic modes that represent scaling are embedded into the multi-frequency POD modes that correspond to scaling of the thermosphere. All the dynamic modes are stable with negligible decay rates. 

\section{Model Limitations}\label{s:ML}
In this section, we discuss the limitations of the quasi-physical dynamic reduced order model for TIE-GCM developed as part of this work. We call it TIE-GCM-ROM-v1.0. The developed model has two major limitations:
\begin{enumerate}
	\item The current version of the model is only applicable for geographic altitudes between 100 and 450 km.
	\item The current version of the model is recommended for use in a $K_{p}$ range from 0 to 5. Application outside of this input range requires further development and/or data assimilation without which the model can result in significantly large errors. 
\end{enumerate}
The model error is on the order of a few percent (3-5\%), which is a function of the number of POD modes used in the reduced order approximation of the dynamic and input matrices (Step 3 of the algorithm described in section \ref{s:NA}).

\section{Further Development}\label{s:FD}
Further development will seek to address the limitations discussed in the previous section while advancing the methodology for modeling the neutral chemical species for a physically self-consistent model and developing the framework for data assimilation. Because the geographic altitude modeled by TIE-GCM is a function of $F_{10.7}$, the extension to higher altitudes will either require extrapolation to higher altitudes under assumptions during low levels of solar activity or development of multiple models valid for smaller ranges of $F_{10.7}$. We will attempt to expand the model's applicability to times to extreme geomagnetic storms. Future work will involve developing the techniques and framework for data assimilation with ROM. 

\section{Conclusions} \label{s:Conc} 
Accurate specification of the thermosphere is important in the context of atmospheric drag, the largest source of uncertainty for orbit prediction in low Earth orbit (LEO), pertinent to space situational awareness. Most existing models can be classified as either empirical (fast to evaluate but with limited forecasting ability) or physics-based (potential for good forecasting abilities but require dedicated parallel resources for real-time application and data assimilative capabilities that have not yet been developed). 

In this work, we develop a quasi-physical dynamic reduced order model (ROM) to overcome the limitations of both empirical and physics-based models. The ROM is developed using modal decomposition methodology, with a newly developed Hermitian Space - Dynamic Mode Decomposition with control (HS-DMDc)  algorithm based on Dynamic Mode Decomposition (DMD). The ROM formulation is also expected to simplify the framework for data assimilation; the development of which will be the subject of future work. We call the developed ROM TIE-GCM-ROM-v1.0. We validate the model by reproducing TIE-GCM output for the years of 2002 and 2008, corresponding to high and low solar activity, respectively, and an independent TIE-GCM simulation in 1996 not used in the development.  Results show that the ROM performs well in serving as a reduced order substitute for TIE-GCM with the ability to maintain low forecast error ($\sim$5\%) for a minimum of 24 hours. The reduced order model evaluation requires very little computational resources with a full day forecast taking only a fraction of a second.

%

\section{acknowledgments}
The first two authors wish to acknowledge support of this work by the Air Force's Office of Scientic Research under Contract Number FA9550-18-1-0149 issued by Erik Blasch. The authors wish to acknowledge useful conversations related to DMD and reduced order modeling with Humberto Godinez of Los Alamos National Laboratory. The authors also wish to thank the anonymous reviewers for their helpful comments. The model can be downloaded at the University of Minnesota Digital Coservancy (https://conservancy.umn.edu/) with a search for TIEGCM-ROM.

\end{document}